\documentclass[times,twocolumn]{aastex63}
\usepackage{amsmath}
\usepackage{amssymb}
\def\kepler{{\textit{Kepler}} }
\newcommand{\au}{~\mbox{AU}}

\newcommand{\bracketfunc}[3]{\left(\frac{#1}{#2}\right)^{#3}}
\newcommand{\rhosurf}{\Sigma}
\newcommand{\vrad}{{v}_{R}}
\newcommand{\vphi}{{v}_{\phi}}
\newcommand{\grav}{{G}}
\newcommand{\mstar}{M_{\ast}}
\newcommand{\mpk}{M_{p,k}}
\newcommand{\rpk}{R_{p,k}}
\newcommand{\mpl}{M_{p}}
\newcommand{\rp}{R_{p}}

\newcommand{\sigmamin}{\rhosurf_{\min}}
\newcommand{\sigmazero}{\rhosurf_0}
\newcommand{\sigmaun}{\rhosurf_{\rm un}}
\newcommand{\sigmaunp}{\rhosurf_{\rm un,p}}

\newcommand{\omegakp}{\Omega_{\rm K,p}}
\newcommand{\omegak}{\Omega_{\rm K}}
\newcommand{\tmig}{\tau_{\rm a}}

\newcommand{\tgap}{t_{\rm gap}}

\newcommand{\hp}{h_{\rm p}}

\newcommand{\rpin}{R_{\rm p,in}}
\newcommand{\rpout}{R_{\rm p,out}}
\newcommand{\mpin}{M_{\rm p,in}}
\newcommand{\mpout}{M_{\rm p,out}}
\newcommand{\tmigin}{\tau_{\rm a,in}}
\newcommand{\tmigout}{\tau_{\rm a,out}}
\newcommand{\pin}{{P}_{\rm in}}
\newcommand{\pout}{{P}_{\rm out}}
\newcommand{\omegain}{\Omega_{\rm K,in}}
\newcommand{\omegaout}{\Omega_{\rm K,out}}
\newcommand{\mptrans}{M_{\rm p,trans}}

\defcitealias{PaperI}{Paper~I}
\defcitealias{Baruteau_Papaloizou2013}{BP13}

\received{}
\revised{}
\accepted{\today}
\submitjournal{ApJ}

\shorttitle{Radial migration of gap-opening planets in protoplanetary disks. II.}
\shortauthors{Kanagawa \& Szuszkiewicz}

\begin{document}

\title{Radial migration of gap-opening planets in protoplanetary disks. II. The case of a planet pair}

\correspondingauthor{Kazuhiro D. Kanagawa}
\email{kazuhiro.kanagawa@utap.phys.s.u-tokyo.ac.jp}

\author[0000-0001-7235-2417]{Kazuhiro D. Kanagawa}
\affiliation{Research Center for the Early Universe, Graduate School of Science, The University of Tokyo, Hongo, Bunkyo-ku, Tokyo 113-0033, Japan}
\affiliation{Institute of Physics and CASA$^{\ast}$, University of Szczecin, Wielkopolska 15, PL-70-451 Szczecin, Poland}

\author[0000-0002-7881-2805]{Ewa Szuszkiewicz}
\affiliation{Institute of Physics and CASA$^{\ast}$, University of Szczecin, Wielkopolska 15, PL-70-451 Szczecin, Poland}




\begin{abstract}
When two planets are born in a protoplanetary disk, they may enter into a mean-motion resonance
as a consequence of the convergent planetary migration. 
The formation of mean-motion resonances is important for understanding how the planetary systems are shaped in the disk environments. 
Motivated by recent progress in the comprehension of the migration of partial gap-opening planets, we have investigated the orbital evolution of the planet pairs in a wide range of masses and disk properties with the aim to find out when the resonance capture is likely to happen.
Using the formula for the migration timescale of a gap-opening planet developed in our previous work, we have derived a simple criterion that allows us to predict when the migration will be convergent (divergent).
Further, we have verified the criterion using two-dimensional hydrodynamic simulations.
We have found that the resonant pair of planets formed at the early phase of evolution can depart from the resonance at later times because the migration speed of the outer planet slows down due to the gap formation.
Moreover, adopting our formula of the migration timescale, we have also carried out three-body simulations, which confirm the results of hydrodynamic simulations.
Finally, we have compared our predictions with the observations, selecting a sample of known two-planet systems.
\end{abstract}


\section{Introduction} \label{sec:introduction}
Planets are born in the protoplanetary gaseous disks.
Their gravitational interaction with the surrounding gas results in the radial orbital migration of planets within the disk \citep[e.g.,][]{Lin_Papaloizou1979,Goldreich_Tremaine1980}.
When a planet is small enough, its migration is described by the linear theory and is known as type~I migration \citep[e.g.,][]{Tanaka_Takeuchi_Ward2002,Paardekooper_Baruteau_Crida_Kley2010}.
A large planet is able to form a density gap along with its orbit, due to strong gravitational disk--planet interactions.
Its motion deviates from the type~I migration due to the gap formation \citep[e.g.,][]{Lin_Papaloizou1986b,Nelson_Papaloizou_Masset_Kley2000,Crida_Morbidelli2007,Edgar2007,
Duffell_Haiman_MacFadyen_DOrazio_Farris2014,Durmann_Kley2015,Dong_Dawson2016,Durmann_Kley2017,PaperI}.

In the ideal case, in which the gap is very deep and no gas is able to pass through the gap, the planet migrates with the viscous drift rate, which is referred to as type~II migration \citep[e.g.,][]{Lin_Papaloizou1986b,Armitage2007}.
However, recent hydrodynamic simulations 
\citep[e.g.,][]{Duffell_Haiman_MacFadyen_DOrazio_Farris2014,Durmann_Kley2015,Durmann_Kley2017} have shown that 
in general, the gap-opening planet is not locked into the viscous evolution of the gas, which means that the migration of such a planet differs from the ``classical'' picture of type~II migration.
For this reason, \cite{PaperI} (hereafter \citetalias{PaperI}) have carried out hydrodynamic simulations for various planet masses, disk aspect ratios, and viscosities, and found that the torque exerted on the planet is roughly proportional to the gas surface density at the bottom of the gap.
This fact indicates that the migration of the gap-opening planet simply slows down due to the reduction in the amount of gas in the vicinity of the planet in the process of gap formation.
In \citetalias{PaperI}, we provide the empirical formula for the migration speed, which can be applied for both the small planet migrating in the linear regime and the gap-opening planet. 
Our formula is able to reproduce reasonably well the migration speed given by hydrodynamic simulations done by us and the authors of the previous studies \citep[i.e.,][]{Duffell_Haiman_MacFadyen_DOrazio_Farris2014,Durmann_Kley2015}.

The gravitational interaction of two planets with the disk and with each other may lead to a mean-motion resonance capture 
\citep[e.g.,][]{Kley_Peitz_Bryden2004,Papaloizou_Szuszkiewicz2005,Quillen2006,Raymond_Barnes_Armitage2008,Rein2012,Ogihara_Kobayashi2013,Migaszewski2015}.
In the mean-motion resonance, the orbital periods of the planets are related to each other as the ratio of two small integers, namely $\pout/\pin= (p+q)/p$, where $\pin$ and $\pout$ are the orbital periods of the inner and outer planets, respectively, and $p$ and $q$ are small integers.
For instance, the planetary system of Gliese~876 is one of the most studied system, containing planets in a 1:2:4 
Laplace resonance \citep{Marcy2001,Rivera2010}.
Many pairs of extra-solar planets in the mean-motion resonance have been confirmed in extra-solar planetary systems \citep[e.g.,][]{Vogt2005,Lee2006,Correia2009,Robertson2012,Giguere2015,Godziewski_Migaszewski_Panichi_Szuszkiewicz2016,Trifonov2017,Migaszewski_Godziewski_Panichi2017}.
On the other hand, \kepler mission has discovered a number of multiple planetary systems composed of close-in super-Earths.
The period ratios of the planet pairs observed in these systems are broadly distributed.
This distribution is overall smooth, but it shows some particular features around some values of the period ratio as e.g., 1.5 and 2.0 \citep{Lissauer2011,Fabrycky2014}.
The interpretation of these features is under investigation and requires taking into consideration both the evolution before and after the disk dispersal.

The formation of a planetary system composed of close-in small planets has been investigated in many previous studies using N-body simulations with the incorporated dissipative forces due to the disk--planet interaction incorporated.
The broad distribution of the period ratio given by the observations can be reproduced by the dynamical instability taking place after the depletion of the gas in the disk \citep[e.g.,][]{Matsumoto_Nagasawa_Ida2012, Hansen_Murray2013,Cossou_Raymond_Hersant_Pierens2014,Ogihara_Morbidelli_Guillot2015,Izidoro_etal2017,Ogihara_Kokubo_Suzuki_Morbidelli2018}.
However, these previous studies used a migration formula for planets that do not open a gap given by, e.g., \cite{Tanaka_Takeuchi_Ward2002} and \cite{Paardekooper_Baruteau_Crida_Kley2010}.
The onset of the dynamical instability is closely connected to the configuration of the planetary system, which is the outcome of the planetary migration within the gaseous disk.
Hence, a realistic model of the planetary migration is essential for those studies.
It indicates that when considering the formation of the close-in planets, we have to take into account the effects of the gap formation on the migration, even if the planet mass is as small as that of a super-Earth.
The systematic survey of hydrodynamic simulations in a broad parameter range done in \citetalias{PaperI} has revealed that the mass of a gap-forming planet becomes smaller as the disk aspect ratio decreases.

\cite{Baruteau_Papaloizou2013} (hereafter \citetalias{Baruteau_Papaloizou2013}) have reported the intriguing evolution of planetary pairs consisting of two shallow gap-forming planets. 
They have shown that even if the planet pair migrates with its period ratio decreasing in time (convergent evolution) and capture into mean-motion resonance happens, the planet pair can depart from the resonant position during
further evolution in the disk.
While the planet pair is leaving resonance, its period ratio increases with time (divergent evolution).
The conditions when the transition from the convergent to the divergent evolution takes place are not fully understood, especially because there might be more than one mechanism responsible for this effect as discussed in \citetalias{Baruteau_Papaloizou2013}.
Here we investigate this problem further by taking advantage of our most recent results on the migration of a single gap-forming planet.
The formula provided in \citetalias{PaperI} appears to be very helpful in understanding the attainment and maintenance of orbital resonances by migrating shallow gap-forming planets.

In this paper, we examine the migration of the planet pairs and the formation of mean-motion resonances, using our simple empirical formula and verify the results by two-dimensional hydrodynamic simulations and three-body simulations.
In Section~\ref{sec:mig_single}, we briefly summarize the result in \citetalias{PaperI} -- the formula for the migration timescale of a single gap-opening planet.
Moreover, in the same section, we describe the prediction of the orbital migration of the planet pair from our formula for the migration timescale of a single gap-opening planet.
Actually, this prediction agrees reasonably well with the results of our hydrodynamic simulations.
In Section~\ref{sec:hydro}, we describe the setup of our hydrodynamic simulations.
Then, we present the results of the hydrodynamic simulations of the planet pair evolution in the disk and discuss the condition for the divergent evolution, comparing the outcome of the simulations with the prediction from our formula for a single gap-opening planet.
In Section~\ref{sec:3body_sims}, we describe a method for the three-body simulations implementing our formula for the migration provided in \citetalias{PaperI}.
In the same section, we present the typical cases of the three-body simulations and the results of our survey in a broad range of masses of the planets.
Section~\ref{sec:discussion} contains a discussion, as well as a comparison with observations, and in Section~\ref{sec:summay} we summarize our results.

\section{Empirical model for the gap-opening planet} \label{sec:mig_single}
\subsection{Formula for a single planet}
In this section, we briefly summarize the results of \citetalias{PaperI} and present our empirical formula for the migration timescale of a single gap-opening planet.
In \citetalias{PaperI}, we have found that the torque exerted on the planet in steady state is roughly proportional to the gas surface density within the bottom of the gap.
According to the previous studies \citep[e.g.,][]{Duffell_MacFadyen2013,Fung_Shi_Chiang2014,Kanagawa2015b}, 
the surface density within the bottom of the gap is given by
\begin{align}
\frac{\sigmamin}{\sigmaunp} &= \frac{1}{1+0.04K}, \label{eq:smin}\\
K&=\bracketfunc{\mpl}{\mstar}{2}\bracketfunc{\hp}{\rp}{-5}\frac{1}{\alpha}, \label{eq:k}
\end{align}
where $\mpl$ and $\mstar$ are the masses of the planet and the central star, respectively, and $\rp$ is the orbital radius of the planet.
The disk scale height is represented by $h$, and $\omegak$ and $\sigmaun$ indicate the Keplerian angular velocity and the unperturbed surface density, respectively.
Hence, the results of our hydrodynamic simulations performed in \citetalias{PaperI} indicate that the torque exerted on the planet in steady state can be given by $\sim (\sigmamin/\sigmaunp)\Gamma_0(\rp)$, where $\Gamma_0$ is defined by
\begin{align}
\Gamma_0(R) &= \bracketfunc{\mpl}{\mstar}{2} \bracketfunc{h}{R}{-2} \sigmaun R^4 \omegak^2 \label{eq:gamma0}.
\end{align}
On the basis of these results, we have given in \citetalias{PaperI} the empirical formula of the migration timescale in steady state in the form  
\begin{align}
\tmig &= -\frac{1+0.04K}{\gamma_L+\gamma_c \exp\left(-K/K_{\rm t} \right)} \tau_0 (\rp),
\label{eq:tmig}
\end{align}
where $\tau_0$ is defined by
\begin{align}
\tau_0 (R)&= \frac{R^2\omegak \mpl}{2\Gamma_0}.
 \label{eq:tau0}
\end{align}
The Lindblad and corotation torques normalized by $\Gamma_0$ are represented by $\gamma_L$ and $\gamma_C$.
In the locally isothermal case, $\gamma_L$ and $\gamma_C$ are described by \citep{Paardekooper_Baruteau_Crida_Kley2010},
\begin{align}
\gamma_L&= - \left(2.5-0.1s+1.7\beta \right) b^{0.71},\label{eq:gamma_L}\\
\gamma_C&= 1.1 \left(1.5-s \right) b + 2.2 \beta b^{0.71} - 1.4\beta b^{1.26}, \label{eq:gamma_C}
\end{align}
where $b=0.4\hp/\epsilon$ and $\epsilon$ is the softening length for the planetary gravitational potential (see below Equation~(\ref{eq:gravpot})), and $s \equiv - d\ln \sigmaun/ d\ln R$ and $\beta \equiv -d\ln T/d\ln R$.
We define the flaring index in the form $f \equiv d\ln (h/R)/d\ln R$, which is related to $\beta$ as follows: $\beta = -2f+1$.
In this paper, we will consider mostly the case of the disks with a constant aspect ratio, for which $f=0$. 
The value of $K_{\rm t}$ is related to the gap depth for which the corotation torque is ineffective and may be taken to be equal to $20$ (see Section~5.1 of \citetalias{PaperI}).

When the planet is small, which means $0.04K \ll 1$ (or $K \ll 25$), the migration timescale is inversely proportional to the mass of the planet.
On the other hand, when the planet is large and $K \gg 25$, the migration timescale in steady state is proportional to the mass of the planet.
Hence, if for simplicity, we ignore in Equation~(\ref{eq:tmig}) the effect of corotation cutoff (which is related to $K_{\rm t}$): thus, the migration timescale is the shortest when the planet mass is equal to $\mptrans$ given as follows:
\begin{align}
\frac{\mptrans}{\mstar} &= 8\times 10^{-5} \bracketfunc{\alpha}{10^{-3}}{1/2} \bracketfunc{\hp/\rp}{0.05}{5/2}.
\label{eq:qtrans}
\end{align}
When the planet mass is larger than $\mptrans$, the migration of the planet slows down due to the gap formation in such a way that for more massive planets the migration is slower.
It is worth noting that in the inner region of the disk, $M_{\rm p,trans}$ is small due to a small disk aspect ratio.
For instance, assuming $\mstar=1M_{\odot}$ and $h/R=0.03$ at around $R=1\au$, we obtain $M_{\rm p,trans} = 7.4 (\alpha/10^{-3})^{1/2} M_{\oplus}$.
This example shows clearly that for a relatively small value of $\alpha$ (e.g., $10^{-3}$), the gap can affect the planetary migration, even if the mass of the planet is of the size of a super-Earth.

Before the gap structure becomes stationary, the migration of the planet can be faster than the saturated value given 
by Equation~(\ref{eq:tmig}).
In \citetalias{PaperI}, using the simple model of the exponential time variation, we have provided the formula taking into account 
the time variation as 
\begin{align}
\tmig &= -\frac{1+0.04K\left[1-\exp\left(-t/\tgap\right)\right]}{\gamma_L+\gamma_c \exp\left(-K/K_{\rm t} \right)} \tau_0 (\rp),
\label{eq:tmig_timevar}
\end{align}
where the gap-opening timescale ($\tgap$) may be given as follows:
\begin{align}
\tgap &= 2.4\times 10^{3} \bracketfunc{\mpl/\mstar}{10^{-3}}{} \bracketfunc{\hp/\rp}{0.05}{-7/2} \bracketfunc{\alpha}{10^{-3}}{-3/2} t_0,
\label{eq:gap_opening_time}
\end{align}
where $t_0 = 2\pi/\omegakp$.
Equation~(\ref{eq:tmig_timevar}) gives a rough fit to the time variation of the migration speed of a single planet obtained in the hydrodynamic simulations, before it reaches the steady-state value (see Section~5.2 of \citetalias{PaperI}).

\subsection{Prediction of the orbital evolution of a planet pair} \label{subsec:prediction_paperI}
Here we consider the radial migration of a planet pair in a protoplanetary disk.
In the following, the subscripts 'in' and 'out' indicate the values of the inner and outer planets, respectively.
The time variation of the period ratio of the planet pair ($\pout/\pin$) is described by
\begin{align}
\frac{\partial}{\partial t} \left(\frac{\pout}{\pin} \right) &= \frac{3}{2} \frac{\omegain}{\omegaout} \left( \tmigin^{-1} - \tmigout^{-1} \right).
\label{eq:timevar_period_ratio}
\end{align}
When the period ratio of the planet pair decreases with time (convergent evolution), it is obvious that, if the planets migrate inwards, the migration timescale of the inner planet is longer than that of the outer planet, namely $\tmigin > \tmigout$.
On the other hand, when the value of $\pout/\pin$ increases with time (divergent evolution), then $\tmigin< \tmigout$.

Note that for the steady-state viscous accretion disks with a constant $\hp/\rp$, the surface density of the unperturbed disk is given by $\Sigma \propto R^{-1/2}$ and the migration timescale is independent of $R$.
This means that the ratio of the migration timescales is also independent of $R$.
However, because the ratio of the migration timescales depends on $R$ in general, we specify the planet locations whenever it is relevant.

\begin{figure}
	\begin{center}
		\resizebox{0.49\textwidth}{!}{\includegraphics{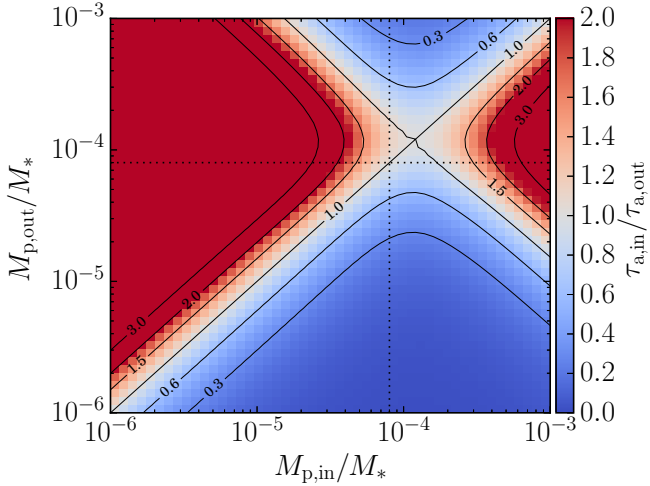}}
		\caption{
		The ratio of $\tmigin$ to $\tmigout$ obtained from Equation~(\ref{eq:tmig}), when $\hp= 0.05$, and $\alpha=10^{-3}$ (which are assumed to be constant throughout the disk).
		We set $\rpin=1$ and $\rpout=1.6$ in the figure.
		The vertical and horizontal dotted lines denote the $\mptrans$ given by Equation~(\ref{eq:qtrans}).
		If $\tmigin/\tmigout > 1$, the evolution of the period ratio of the planet pair would be convergent, while it would be divergent if $\tmigin/\tmigout < 1$.
		\label{fig:tmig_ratio_a1e-3_h5e-2}
		}
	\end{center}
\end{figure}
In this subsection, we consider the case where two planets stay close to 2:1 mean-motion resonance, one of the configurations of interest.
Thus, $\rpin=R_0$ and $\rpout=1.6R_0$ in the rest of this section.
In Figure~\ref{fig:tmig_ratio_a1e-3_h5e-2}, we present the values of $\tmigin/\tmigout$ calculated from Equation~(\ref{eq:tmig}) for the most interesting ranges of masses of the inner and outer planets, when $\alpha=10^{-3}$ and $\hp/\rp =0.05$ are constant throughout the disk. 
The particular role in making the prediction for the ratio of the migration timescales plays the value of the particular mass, namely $\mptrans$, which tell us what is the value of mass of the planet where the transition from type~I migration to the gap-opening planet type migration takes place. 
The whole plane defined by the inner planet mass versus outer planet mass can be divided into four domains using two lines of $\mpin=\mpout$ and $\mpin/\mptrans = \mptrans/\mpout$.
Along these lines, $\tmigin = \tmigout$.

When the masses of both planets in the pair are smaller than $\mptrans$ (which is around $10^{-4}$ in the case presented in Figure~\ref{fig:tmig_ratio_a1e-3_h5e-2}), the migration of the planets can be described by the type~I formula.
It is evident from Figure~\ref{fig:tmig_ratio_a1e-3_h5e-2} that the simple expression in Equation~(\ref{eq:qtrans}) provides a good approximation to the actual value of $\mptrans$.
In this case, only when $\mpout > \mpin$, the outer planet can catch up with the inner planet, and thus the planet pair can be locked in the resonance.
When $\mpin>\mptrans$ but $\mpout<\mptrans$, only the migration of the inner planet slows down, whereas the outer planet migrates in the type~I regime.
Hence, the planet pair can enter the resonance when $(\mpin/\mptrans) > (\mptrans/\mpout)$.
Similarly, when $\mpin<\mptrans$, the planets in the pair can be captured into resonance if $(\mpin/\mptrans) < (\mptrans/\mpout)$.
When the masses of both the inner and outer planets are larger than $\mptrans$, the migration of the planets slows down according to their mass, the more massive planets are slowed down more.
Hence, the migration of the outer planet is faster than that of the inner planet when $\mpin>\mpout$.

\begin{figure}
	\begin{center}
		\resizebox{0.49\textwidth}{!}{\includegraphics{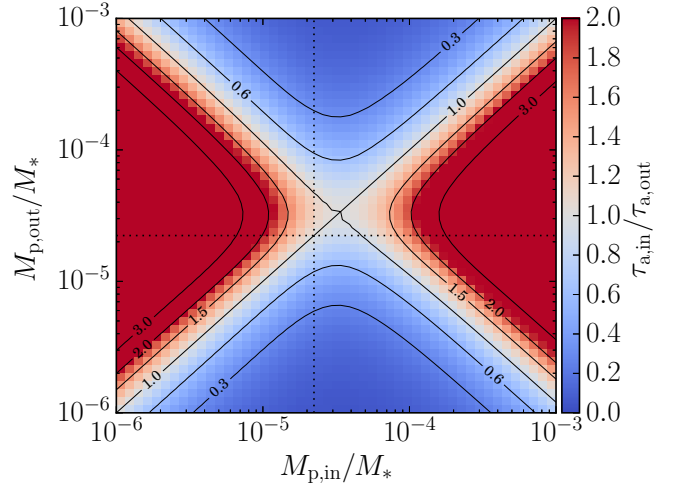}}
		\caption{
		The same as Figure~\ref{fig:tmig_ratio_a1e-3_h5e-2}, but for the case with $h/R=0.03$.
		\label{fig:tmig_ratio_a1e-3_h3e-2}
		}
	\end{center}
\end{figure}
Observations of exoplanets have revealed a number of close-in super-Earths.
These planets would experience the radial migration in the inner region of the protoplanetary disk at the early phases of the evolution.
As mentioned above, the migration of the low-mass planets can slow down due to the gap formation (see Equation~(\ref{eq:qtrans})) when the disk aspect ratio is small.
In the case of $\hp/\rp=0.03$ and $\alpha=10^{-3}$, Figure~\ref{fig:tmig_ratio_a1e-3_h3e-2} illustrates $\tmigin/\tmigout$ calculated from Equation~(\ref{eq:tmig}).
The overall picture is almost the same as in Figure~\ref{fig:tmig_ratio_a1e-3_h5e-2}, but the range of masses is scaled down to the mass range including super-Earths.

We should note that Equation~(\ref{eq:tmig}) can reproduce the migration speed of the planets obtained in the hydrodynamic simulations, with the accuracy within a factor of 2 -- 3 (see Figure~8 in \citetalias{PaperI}).
As a consequence, the above prediction would not be accurate when $\tmigin \sim \tmigout$.
Nonetheless, when $\tmigin \gg \tmigout$ or $\tmigin \ll \tmigout$, the above discussion is useful to understand the orbital migration of the planet pairs.

\section{Hydrodynamic simulations} \label{sec:hydro}
As discussed in the previous section, we may be able to predict the orbital evolution of the planet pairs, using Equation~(\ref{eq:tmig}).
In this section, we carry out hydrodynamic simulations of the planet pairs in order to confirm the validity of the prediction made in the previous section.

\subsection{A brief summary of our computational setup} \label{subsec:method}
We use the two-dimensional numerical hydrodynamic code {\sc \tt FARGO} \footnote{See  \url{http://fargo.in2p3.fr/}} \citep{Masset2000} to simulate the evolution of two planets in a protoplanetary disk.
Basically, the numerical method and setup are the same as in \citetalias{PaperI}, except that now there are two planets in the disk instead of one.
Here, we briefly summarize the setup of our hydrodynamic simulations.

We assume a geometrically thin and non-self-gravitating disk.
We use a two-dimensional cylindrical coordinate system ($R,\phi$), and its origin is located at the position of the central star.
The surface density is represented by $\Sigma$ and the velocities in the radial and azimuthal directions are denoted by ($\vrad,\vphi$).
We adopt a simple, locally isothermal equation of state, and the temperature does not depend on time.

The gravitational potential $\Psi$ is given by
\begin{align}
\Psi =& -\frac{\grav \mstar}{R}\nonumber \\
&-\sum_{k}^{2} \left[ \frac{\grav \mpk}{\left[ R^2+2R\rpk \cos\left( \phi-\phi_{p,k} \right) + \rpk^2 +\epsilon^2 \right]^{1/2}} \right] \nonumber \\
&+\sum_{k}^{2} \left[ \frac{\grav \mpk}{\rpk^2} R\cos\left( \phi-\phi_{p,k} \right) \right],
	\label{eq:gravpot}
\end{align}
where $\grav$ is the gravitational constant, $\mstar$ is the mass of the central star, and $\mpk$ is the mass of the inner ($k=1$) and the outer ($k=2$) planets, which are located at ($\rpk$,$\phi_{p,k}$), respectively.
The softening length is denoted by $\epsilon$.
The first and second terms in equation~(\ref{eq:gravpot}) are the gravitational potentials of the planet and the central star, respectively.
The third term is an indirect term that reflects the fact that the coordinate system based on the central star is not inertial.

The softening length $\epsilon$ in the gravitational potential given in equation~(\ref{eq:gravpot}) is set to $0.6$ times the disk scale height at the location of the planet.
Considering the existence of the circumplanetary disk, we exclude $60\%$ of the planets' Hill radius when calculating the force exerted by the disk 
on the planet, following \citetalias{Baruteau_Papaloizou2013}.
We use an arbitrary value $R_0$ as a unit of distance and $\mstar$ (the mass of the central star) as a unit of mass.
The masses of the central star and the planet pair are assumed to be independent of time, for simplicity.
In our fiducial setting, the initial orbital radii of the planets are $R_0$ for the inner and $1.7R_0$ for the outer planet, respectively.
The computational domain is divided equally into $1024$ meshes in the radial direction and into $2048$ meshes in the azimuthal direction.
Considering the viscous accretion disk to be in steady state, we assume the initial surface density distribution to be $\Sigma=\Sigma_0 (R/R_0)^{-1/2}$.
The value of $\alpha$ is constant throughout the disk.
We adopt the so-called ``open'' boundary condition in the inner boundary, and at the outer boundary, the physical quantities are fixed on the initial 
values during the simulations.
In addition, the wave-killing zones are set near the inner and outer boundaries (for details, see \citetalias{PaperI}).

\begin{table*}
	\begin{center}
 		\caption{Parameters and Results of Runs}
 		\label{tab:models}
 		\begin{tabular}{ccccccccc}
 		\hline \hline
 		Label & $\mpin/\mstar$ & $\mpout/\mstar$ & $\Sigma_0$ & $h/R$ & $\alpha$ &  $\tmigin/\tmigout$\tablenotemark{a} & Evolution Feature \tablenotemark{b}\\
 		\hline
 		Run~1 & $8\times 10^{-5}$ & $3\times 10^{-4}$   & $3\times 10^{-4}$ & $0.05$ & $10^{-3}$ & $0.66$ & convergent $\rightarrow$ divergent \\
 		Run~2 & $8\times 10^{-5}$ & $3\times 10^{-4}$   & $1\times 10^{-4}$ & $0.05$ & $10^{-3}$ & $0.66$ &  convergent $\rightarrow$ divergent \\
 		Run~3 & $8\times 10^{-5}$ & $3\times 10^{-4}$   & $5\times 10^{-4}$ & $0.05$ & $10^{-3}$ & $0.66$ &  convergent $\rightarrow$ divergent\tablenotemark{c}\\
 		Run~4 & $3\times 10^{-4}$ & $5\times 10^{-4}$   & $3\times 10^{-4}$ & $0.05$ & $10^{-3}$ & $0.63$ & divergent \\
 		Run~5 & $5\times 10^{-5}$ & $8\times 10^{-5}$   & $3\times 10^{-4}$ & $0.05$ & $10^{-3}$ & $1.43$ & convergent (3:2 MMR) \\
 		Run~6 & $3\times 10^{-4}$ & $1.5\times 10^{-4}$ & $3\times 10^{-4}$ & $0.05$ & $10^{-3}$ & $1.59$ & divergent $\rightarrow$ convergent\\
 		Run~7 & $5\times 10^{-4}$ & $3\times 10^{-4}$   & $3\times 10^{-4}$ & $0.05$ & $10^{-3}$ & $1.58$ & convergent (3:2 MMR) \tablenotemark{c}\\
 		\hline
 		Run~8 & $2\times 10^{-5}$ & $1\times 10^{-4}$ & $1\times 10^{-4}$ & $0.03$ & $10^{-3}$ & $0.61$ & convergent $\rightarrow$  divergent \\
 		Run~9 & $5\times 10^{-5}$ & $1\times 10^{-4}$ & $1\times 10^{-4}$ & $0.03$ & $10^{-3}$ & $0.59$ & convergent $\rightarrow$  divergent \\
 		Run~10 & $2\times 10^{-5}$ & $5\times 10^{-5}$ & $1\times 10^{-4}$ & $0.03$ & $10^{-3}$ & $1.03$ & convergent (3:2 MMR) \\
 		Run~11 & $1\times 10^{-5}$ & $2\times 10^{-5}$ & $1\times 10^{-4}$ & $0.03$ & $10^{-3}$ & $1.84$ & convergent (3:2 MMR) \\
 		Run~12 & $1\times 10^{-4}$ & $5\times 10^{-5}$ & $1\times 10^{-4}$ & $0.03$ & $10^{-3}$ & $1.70$ & convergent (3:2 MMR) \\
 		\hline
 		Run~13 & $5\times 10^{-5}$ & $2\times 10^{-4}$ & $1\times 10^{-4}$ & $0.03$ & $10^{-2}$ & $1.05$ & convergent $\rightarrow$  divergent \\
 		Run~14 & $5\times 10^{-5}$ & $8\times 10^{-5}$ & $1\times 10^{-4}$ & $0.03$ & $10^{-2}$ & $1.36$ & convergent (2:1 MMR) \\
 		Run~15 & $5\times 10^{-5}$ & $1\times 10^{-4}$ & $1\times 10^{-4}$ & $0.03$ & $10^{-2}$ & $1.42$ & convergent (2:1 MMR) \\
  		\hline
 		\end{tabular}
 	\end{center}
	\tablenotetext{a}{The ratio of the migration timescales calculated from Equation~(\ref{eq:tmig}) when $\rpout/\rpin = 1.6$.}
	\tablenotetext{b}{A feature of the evolution of the planet pair given by our hydrodynamic simulations. The detailed description is in Section~\ref{subsec:param_study}.}
	\tablenotetext{c}{Gaps are (partially) merged.}
\end{table*}
In the following subsections, we present the results of our hydrodynamic simulations.
The masses of the planets, the surface density of the disk, the disk aspect ratio, and the value of $\alpha$ in each run are listed in Table~\ref{tab:models}.
For reference, in the table, we list the ratio of $\tmigin$ to $\tmigout$ calculated from Equation~(\ref{eq:tmig}) where $\rpout/\rpin = 1.6$ and $\rpin=R_0$.
As discussed in Section~\ref{subsec:prediction_paperI}, when $\tmigin/\tmigout < 1$, the evolution of a planet pair is expected to be divergent, while it is expected to be convergent when $\tmigin/\tmigout > 1$. 
In the table, we also summarize the features of each evolution, which is determined from hydrodynamic simulations (for details, see Section~\ref{subsec:param_study}).

\subsection{A reference case} \label{subsec:reference_case}
\subsubsection{The orbital evolution of planets and surface density distribution in the disks}
First, we present the outcomes of hydrodynamic simulations with $\mpin/\mstar=8\times 10^{-5}$ and $\mpout/\mstar=3\times 10^{-4}$ when 
$\alpha=10^{-3}$, $\hp/\rp =0.05$ and $\Sigma_0=3\times 10^{-4}$ (Run~1), as a reference case.
\begin{figure*}
	\begin{center}
		\resizebox{0.98\textwidth}{!}{\includegraphics{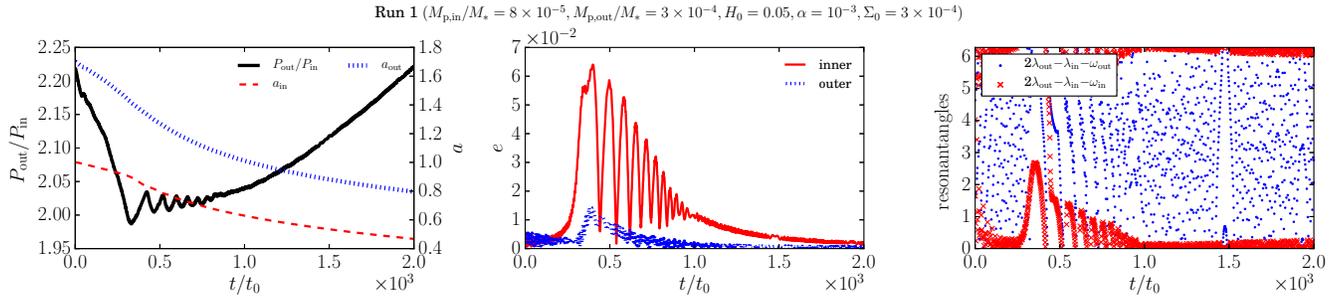}}
		\caption{
		The orbital evolution in the case of Run~1 ($\mpin/\mstar=8\times 10^{-5}$, $\mpl/\mstar=3\times 10^{-4}$ and $h/R=0.05$ and $\alpha=10^{-3}$, $\Sigma_0=3\times 10^{-4}$).
		From the left, the time variations of the semi-major axes of the inner and outer planets ($a_{\rm in}$ and $a_{\rm out}$, respectively), and their period ratio, the time variations of the eccentricities of the inner and outer planets, and the resonant angles relevant to the period ratio of the planet pair are shown, where $\lambda$ and $\omega$ denote the mean longitude and longitude of periastron.
		\label{fig:qin8e-5_qout3e-4_h5e-2_a1e-3_S3e-4}
		}
	\end{center}
\end{figure*}
Figure~\ref{fig:qin8e-5_qout3e-4_h5e-2_a1e-3_S3e-4} shows the time variations of the semi-major axes, the orbital period ratio, the eccentricities, and the resonant angles for the 2:1 mean-motion resonance.
In this case, both planets migrate inward as can be seen in the figure.
At the early phase of the evolution ($t \lesssim 500\ t_0$), the period ratio decreases, and the evolution of the planet pair is convergent.
Around $t=500 \ t_0$, the planet pair enters into 2:1 mean-motion resonance, and as a result of this, the eccentricity of the inner planet increases.
However, after $t\sim 1000\ t_0$, the period ratio starts to increase with time, and the planet pair departs from the resonance position.
As the planet pair leaves from the resonance position, the eccentricity of the inner planet decreases.
The outcome of this simulation is the divergent evolution of the planets, which starts at around $t\sim 1000\ t_0$ and lasts until the end of the calculations. 

\begin{figure}
	\begin{center}
		\resizebox{0.49\textwidth}{!}{\includegraphics{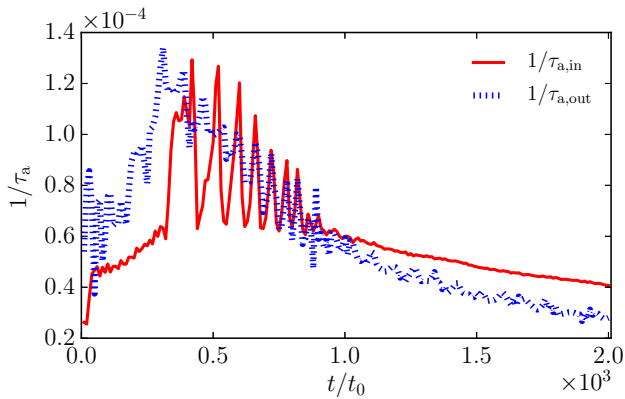}}
		\caption{
		The time variations of the migration timescale in the case of Run~1.
		The plotted migration timescale is averaged during $10 \ t_0$.
		\label{fig:qin8e-5_qout3e-4_h5e-2_a1e-3_S3e-4_migration_speed}
		}
	\end{center}
\end{figure}
Figure~\ref{fig:qin8e-5_qout3e-4_h5e-2_a1e-3_S3e-4_migration_speed} shows the time variations of the migration timescale of the inner and outer planets.
In the early phase of the evolution from the beginning until $t=500\ t_0$, the migration timescale of the outer planet is shorter than that of the inner planet.
After $t=500 \ t_0$, the migration timescale of the outer planet becomes longer, and finally, the migration of the outer planet is slower than that of the inner planet at around $t=1000\ t_0$.
Although the migration timescale of the inner planet is affected by the planet--planet interaction around $t=500\ t_0$, it does not change that much during the whole simulation.
It is reasonable to consider that the transition from the convergent to divergent evolution originates from the slowdown of the migration of the outer planet.

\subsubsection{Effects of planet--wake interactions}
\begin{figure}
	\begin{center}
		\resizebox{0.49\textwidth}{!}{\includegraphics{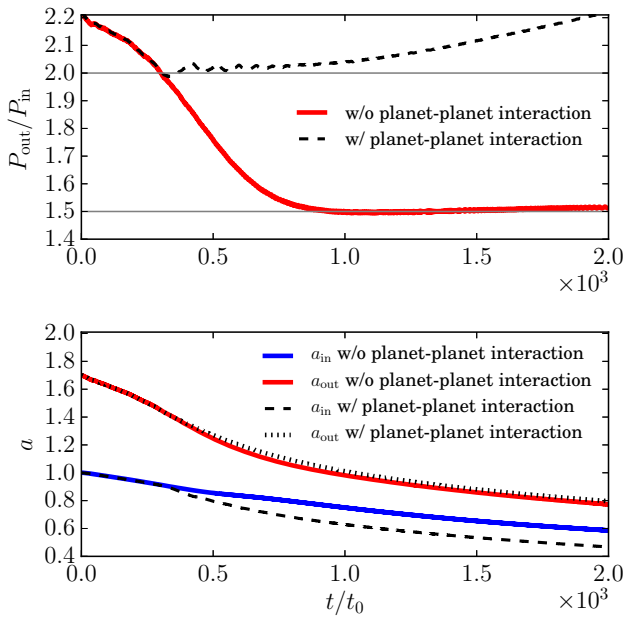}}
		\caption{
		The time variations of the period ratio (upper) and the semi-major axis (lower) in the cases with and without the planet-planet interaction.
		\label{fig:qin8e-5_qout3e-4_h5e-2_a1e-3_S3e-4_period_ratio_comp}
		}
	\end{center}
\end{figure}
\citetalias{Baruteau_Papaloizou2013} reported a few cases of the evolution of a pair of planets in a protoplanetary disk, in which the transition from the convergent to the divergent evolution occurs, in a similar way to what has been shown in Figure~\ref{fig:qin8e-5_qout3e-4_h5e-2_a1e-3_S3e-4}.
They have argued the possibility that this transition is caused by the planet--wake interactions, when the density waves launched by the one of the planets penetrate into the co-orbital region of the other planet.
A convincing argument in favor of this mechanism has been provided by their simulations in which the planet--planet interaction is turned off, and the transition from the convergent to the divergent evolution still takes place. 
In order to identify the reason for the transition in our case, which differs from that of \citetalias{Baruteau_Papaloizou2013} not only in the masses of the planets and the disk parameters, but also in the viscosity formulation (they have assumed that $\nu$ is constant throughout the disk; instead, our assumption is that $\alpha$ is constant), we also carried out a hydrodynamic simulation in which the planet-planet interaction is turned off. 
\footnote{In our simulations, we do not exclude the indirect term, as it is different from that of BP13. However, we confirmed that switching on/off the indirect term does not affect the results significantly.}
In Figure~\ref{fig:qin8e-5_qout3e-4_h5e-2_a1e-3_S3e-4_period_ratio_comp}, we compare the time variations of the period ratio and the semi-major axes of the inner and outer planets when the planet--planet interaction is taken into account with that when it is ignored.
In the case where the planet--planet interaction is ignored, the period ratio monotonically decreases and finally reaches the stationary value of $\sim 1.5$.
On the other hand, in the case in which the planet--planet interaction is considered, the evolution of the planet pair is convergent until the period ratio reaches the value of $\sim 2.0$ and after that, it becomes divergent.
Hence, in the case of Run~1, the planet--planet interaction plays an important role in making the transition from the convergent to the divergent evolutions to happen.
Note that the period ratio in the case without the planet-planet interaction becomes stationary with the value of $1.5$, 
but it is not related to the mean-motion resonance (the 3:2 resonance angles do not librate).

\begin{figure*}
	\begin{center}
		\resizebox{0.49\textwidth}{!}{\includegraphics{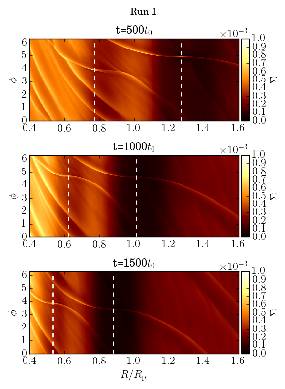}}
		\resizebox{0.49\textwidth}{!}{\includegraphics{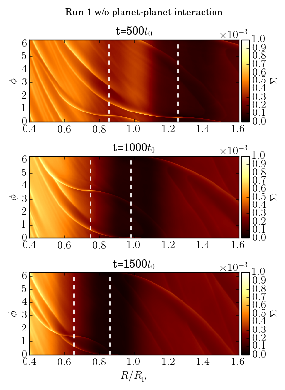}}
		\caption{
		The two-dimensional distributions of the surface density at $t=500\ t_0$, $1000\ t_0$, and $t=1200\ t_0$ from the top, in the case of Run~1.
		In the left panels, the planet-planet interaction is included and in the right panels, it is not included.
		The two vertical dashed lines indicate the orbital radii of the inner and outer planets.
		\label{fig:qin8e-5_qout3e-4_h5e-2_a1e-3_S3e-4_gasdens_comp}
		}
	\end{center}
\end{figure*}
Figure~\ref{fig:qin8e-5_qout3e-4_h5e-2_a1e-3_S3e-4_gasdens_comp} illustrates the two-dimensional distributions of the surface density in two cases: in the left panel, planet--planet interaction is considered, and in the right panel, it is ignored.
In both cases, density waves excited by the one of the planets clearly penetrate into the co-orbital region of the other planet.
In the case where planet--planet interaction is considered, the gaps in the disk formed by the planets are separated from each other.
On the other hand, in the case where planet--planet interaction is ignored, the outer planet arrives closer to the inner planet and two planets form a common gap.

\begin{figure}
	\begin{center}
		\resizebox{0.49\textwidth}{!}{\includegraphics{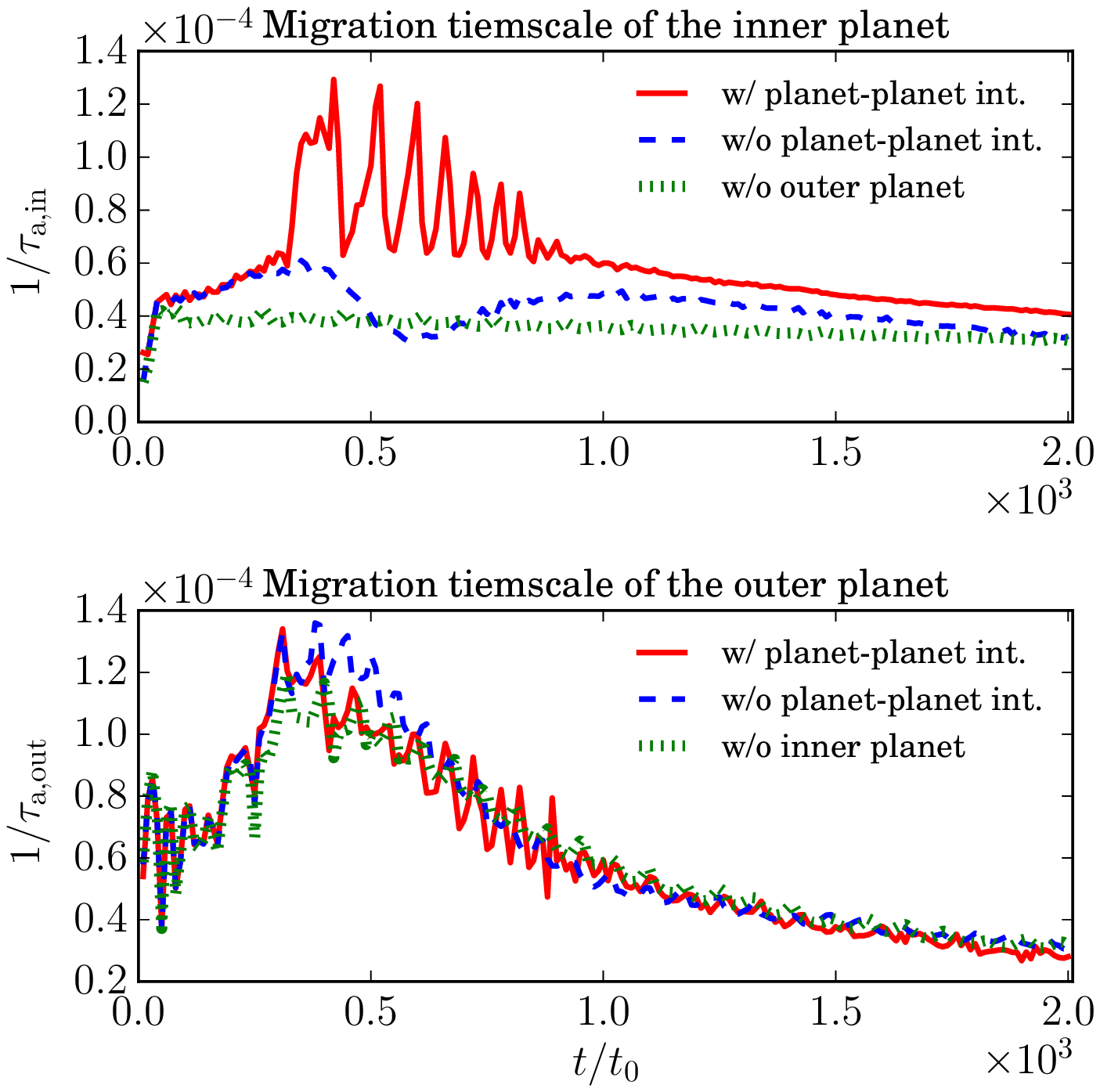}}
		\caption{
		The time variations of the migration timescale of the inner planet (upper) and the outer planet (lower) in the case of Run~1.
		\label{fig:qin8e-5_qout3e-4_h5e-2_a1e-3_S3e-4_migration_speed_comp_wopp}
		}
	\end{center}
\end{figure}
In Figure~\ref{fig:qin8e-5_qout3e-4_h5e-2_a1e-3_S3e-4_migration_speed_comp_wopp}, we compare the migration timescales in the cases with and without the planet-planet interaction.
For reference, we have carried out additional hydrodynamic simulations with one planet (inner or outer planet only), and we plot the migration timescales given by these simulations in the same figure.
As can be seen in Figure~\ref{fig:qin8e-5_qout3e-4_h5e-2_a1e-3_S3e-4_migration_speed_comp_wopp}, the migration timescale of the inner planet is significantly influenced by the outer planet.
In contrast, the migration timescale of the outer planet is hardly affected by the inner planet.
As shown in \citetalias{PaperI}, although the migration speed is fast at the early phase of the evolution, it decreases later on as the gap opens.
The time variation of the migration timescale of the outer planet can be explained by the slowdown of the migration due to the gap formation.
In the case where planet--planet interaction is switched on, one can consider that the transition from the convergent to the divergent evolution is caused by the slowdown of the outer planet migration and the slight speed-up of the inner planet migration due to the planet--planet interaction.
In the case where planet--planet interaction is ignored, 
the outer planet can move close to the inner planet, and the planet pair forms the common gap.
We further discuss the effect of the common gap formation in Appendix~\ref{sec:common_gap}.

\subsection{Parameter study} \label{subsec:param_study}
\begin{figure}
	\begin{center}
		\resizebox{0.49\textwidth}{!}{\includegraphics{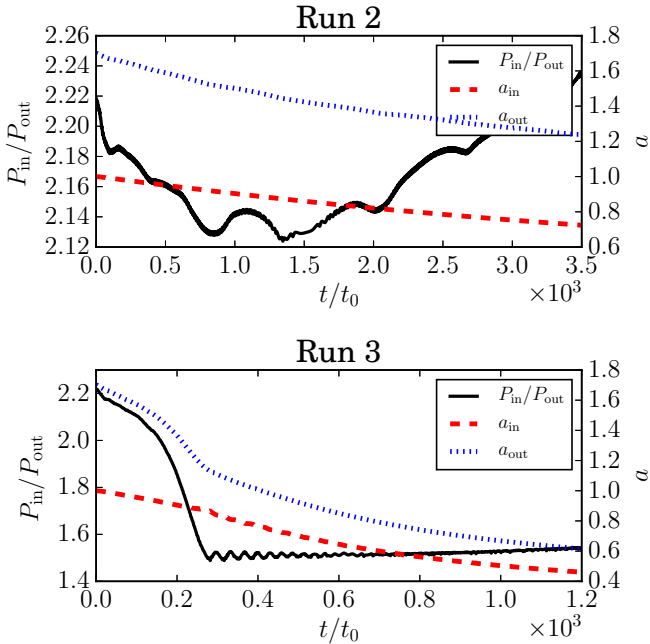}}
		\caption{Evolutions of the period ratio of the planet pair (left axis) and semi-major axes (right axis) in the cases of Run~2 (upper) and Run~3 (lower).
		\label{fig:orbitevo_runs2-3}
		}
	\end{center}
\end{figure}
In this section, we investigate the cases in which the planet masses and disk parameters are different from those in the reference case.
First, in Figure~\ref{fig:orbitevo_runs2-3}, we show the orbital evolutions with the same planet masses ($\mpin/\mstar=8\times 10^{-5}$ and $\mpout/\mstar=3\times 10^{-4}$), and the same values of $\hp/\rp$ and $\alpha$ as in the reference case ($\hp/\rp =0.05$ and $\alpha =10^{-3}$), but where the surface density of the disk gas is smaller (Run~2) and larger (Run~3).
In the cases of Run~2 ($\Sigma_0=1\times 10^{-4}$), the period ratio of the planet pair decreases in the early stage, but the planet pair does not enter any mean-motion resonance.
After $t\simeq 1500 \ t_0$, the period ratio increases with time, and the evolution becomes divergent.
In the case of Run~3 ($\Sigma_0=5 \times 10^{-4}$), the initial inward migration of the outer planet is fast, and hence, the outer planet can be closer to the inner planet.
As a result, the planet pair is captured into the 3:2 mean-motion resonance around $t=300 \ t_0$.
After $t=300 \ t_0$, the period ratio slowly increases with time, and the eccentricities of the planet decrease down to $\sim 10^{-2}$, as the gaps open.
In this case, as different from the cases of Runs~1 and 2, the gaps formed by the planets are partially merged because the separation between the planets is narrower.
Due to the effect of gap merging, the rate of increase of the period ratio is slow, but it can be considered to be divergent evolution.
When the gas surface density is smaller or larger than that in the reference case, the evolution of the planet pair is divergent as in the reference case.
This trend is consistent with the prediction described in Section~\ref{subsec:prediction_paperI}, because the ratio of $\tmigin$ to $\tmigout$ is independent of $\Sigma_0$ when $f=0$ and $s=0.5$.

\begin{figure*}
	\begin{center}
		\resizebox{0.98\textwidth}{!}{\includegraphics{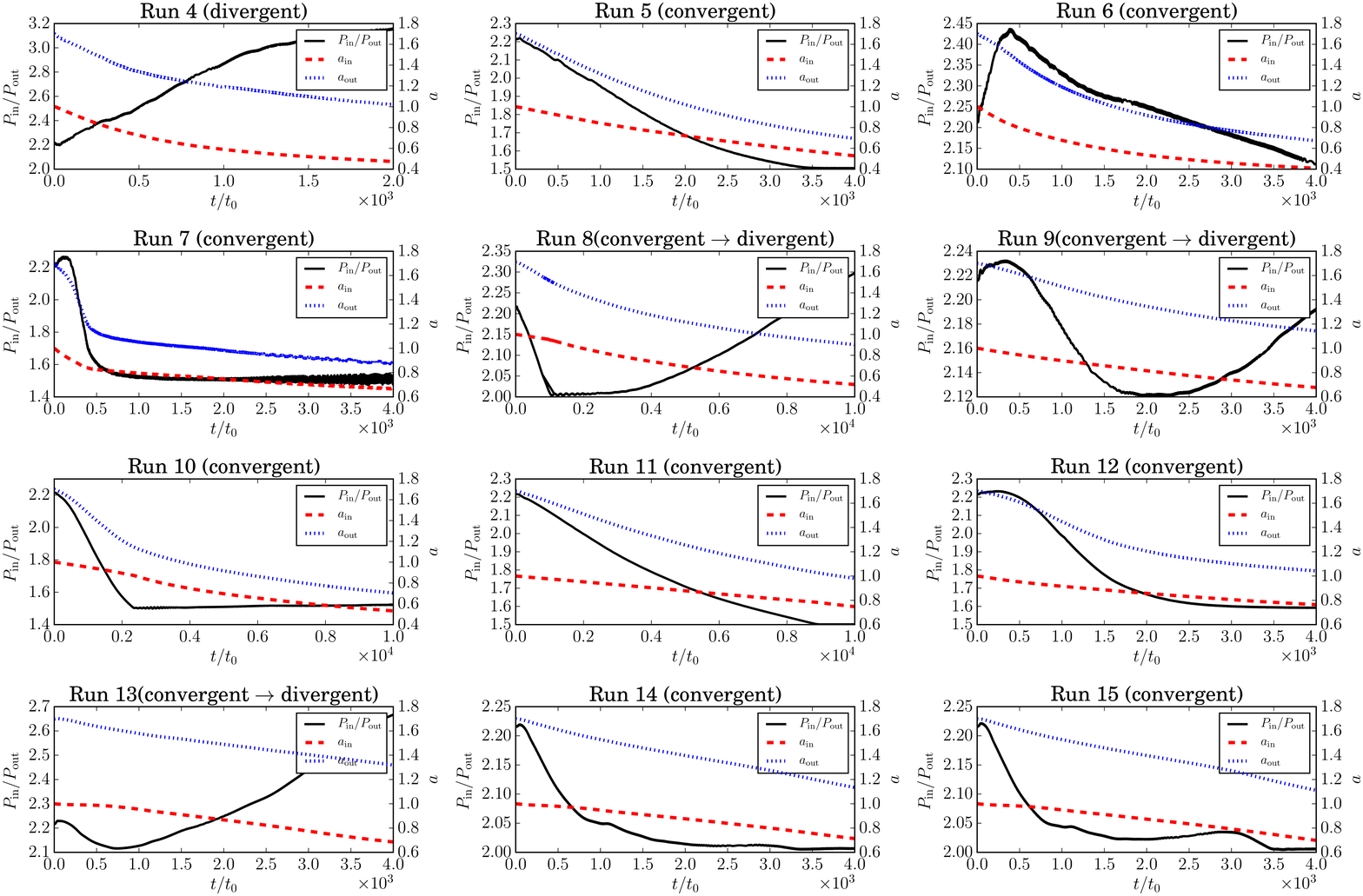}}
		\caption{The same as Figure~\ref{fig:orbitevo_runs2-3}, but for Runs~4 -- 15.
		\label{fig:orbitevo_runs4-15}
		}
	\end{center}
\end{figure*}
Figure~\ref{fig:orbitevo_runs4-15} illustrates the orbital evolutions as shown in Figure~\ref{fig:orbitevo_runs2-3} but for the various masses of the planet pair and the disk parameters ($\alpha$ and $h/R$).
The parameters are listed in Table~\ref{tab:models}.
We simulated the orbital evolutions of the planet pair at least until $4000 \ t_0$ or until the time when the inner planet reaches $R=0.5 R_0$.
In the case of Run~4, the period ratio increases during almost the whole simulation time, and the evolution of the planet pair can be labeled as divergent.
In the case of Run~6, the period ratio decreases with time after $t\simeq 300\ t_0$ and the evolution of the planet pair is convergent, while the period ratio decreases in $t \lesssim 300\ t_0$.
Note that in the case of Run~6, though the evolution is convergent, the planet pair cannot reach the 2:1 mean-motion resonance until the inner planet reaches $R=0.4R_0$.
In Runs~9 and 13, though the evolution of the planet pair is convergent in the early phase, eventually, the evolution becomes divergent.
In the other cases (Runs~5, 7, 10, 11, 12, 14, and 15), the period ratio decreases in the early phase, and eventually the planet pair is captured into the mean-motion resonance.
The evolution features described above are summarized in Table~\ref{tab:models} and we also denote the label on the top of each plot.


\begin{figure}
	\begin{center}
		\resizebox{0.45\textwidth}{!}{\includegraphics{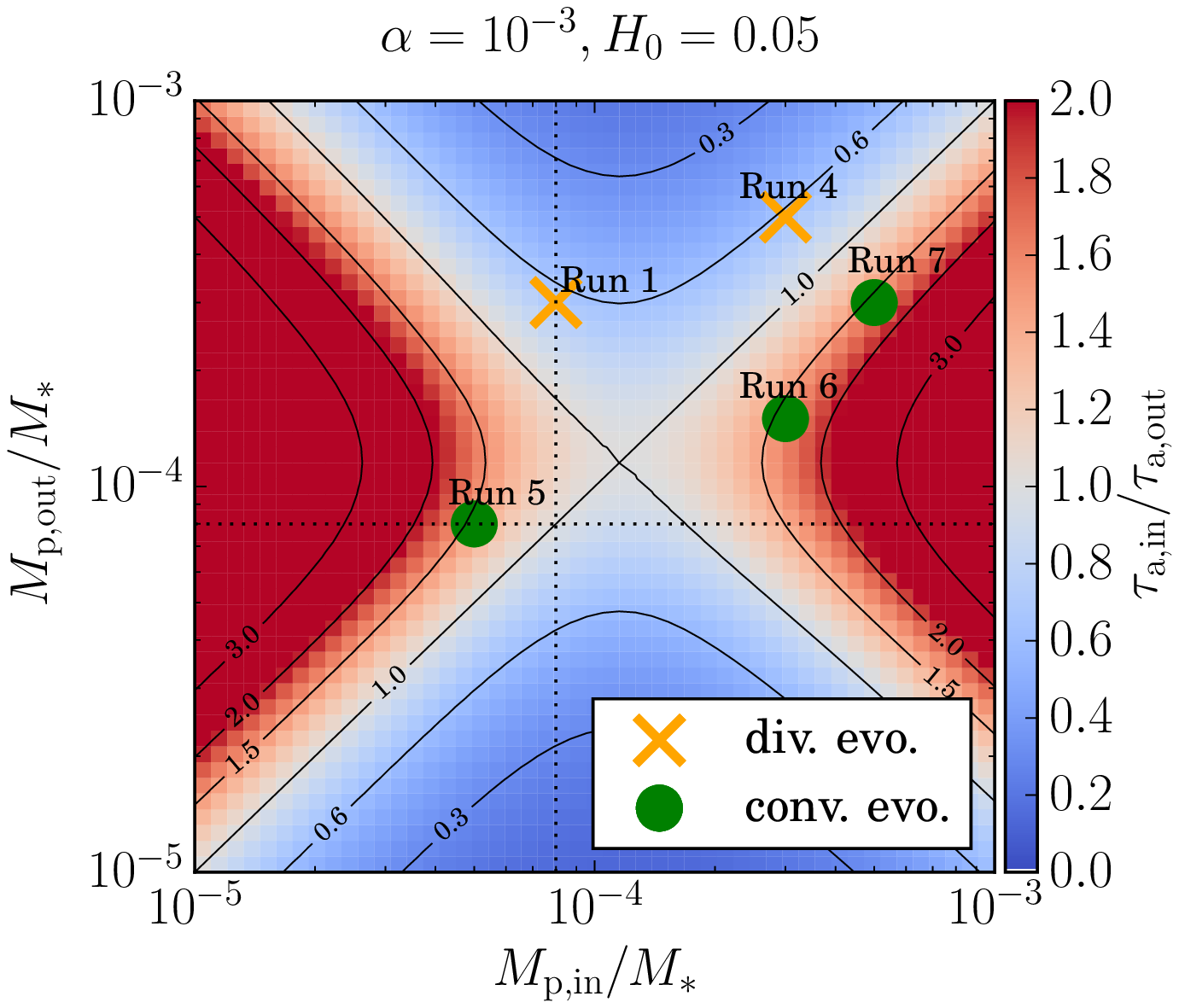}}
		\resizebox{0.45\textwidth}{!}{\includegraphics{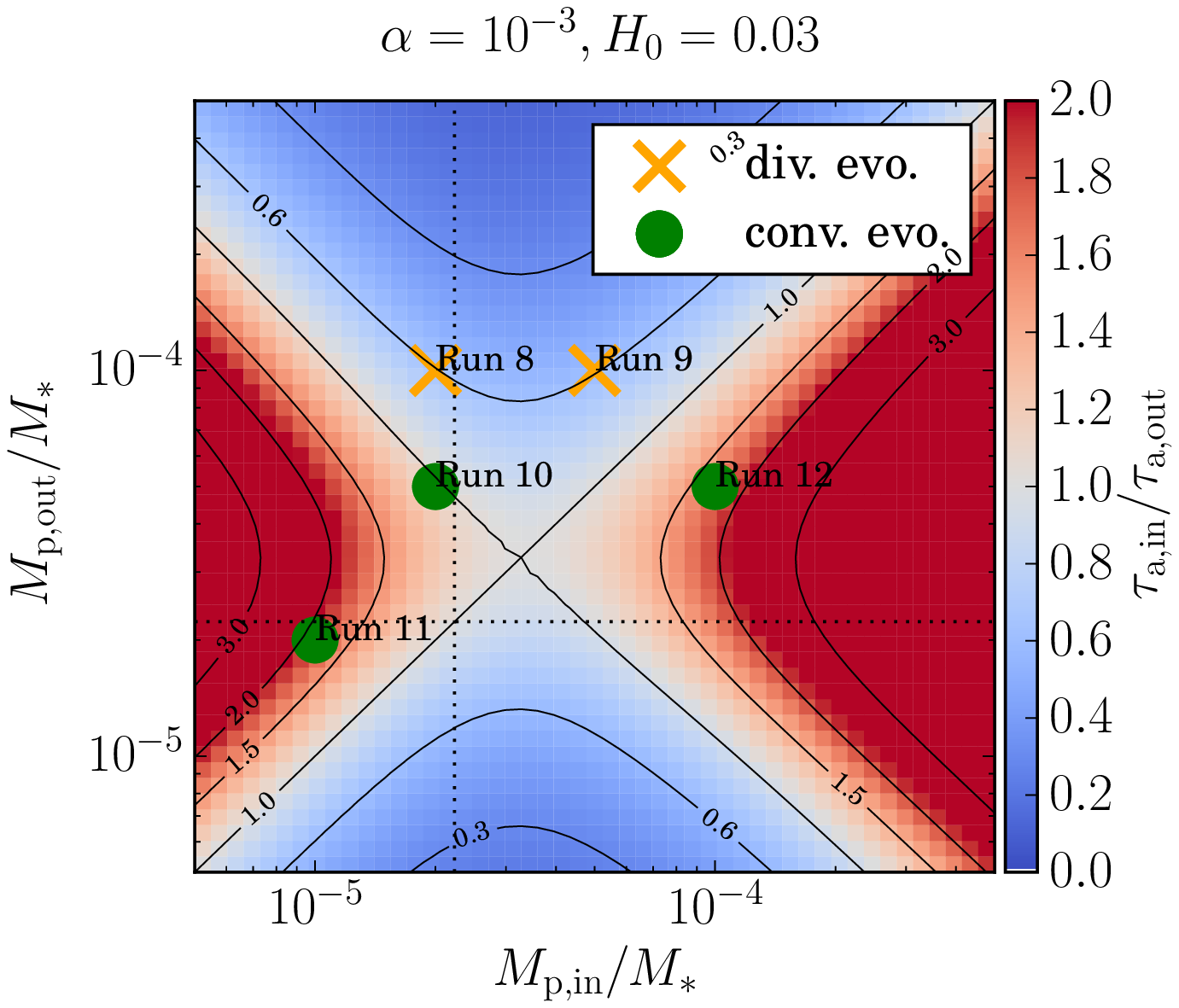}}
		\resizebox{0.45\textwidth}{!}{\includegraphics{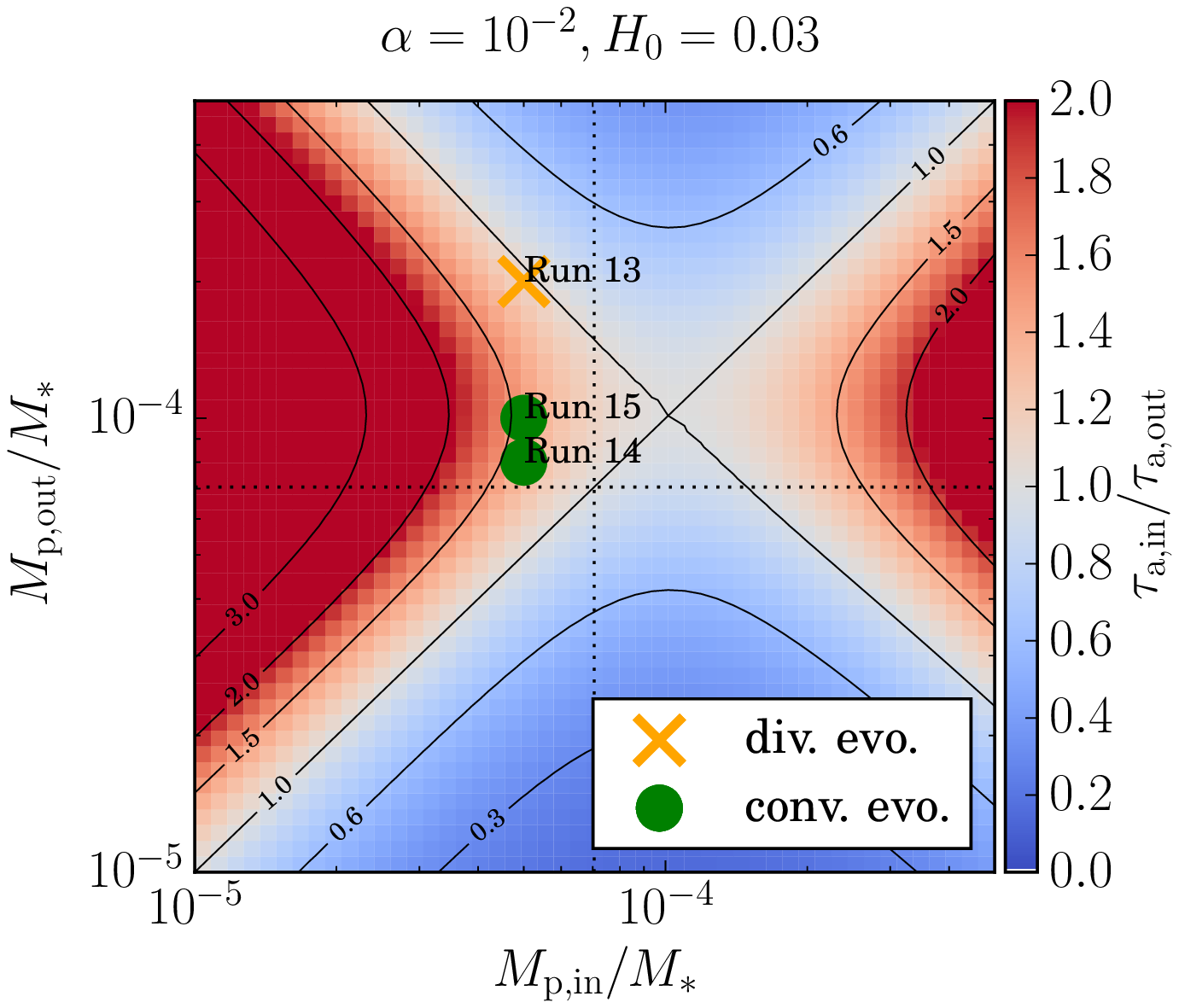}}
		\caption{
		The types of the orbital evolution of the planet pair given by our simulations, plotted over the map of the ratio of $\tmigin$ to $\tmigout$ calculated by Equation~(\ref{eq:tmig}) for $\alpha=10^{-3}, h/R=0.05$ (upper) (for Runs 1, 4 -- 7), $\alpha=10^{-3}, h/R=0.03$ (middle) (for Runs~8 -- 12), and $\alpha=10^{-2},h/R=0.03$ (lower) (for Runs~13--15).
		The cross and circle indicate divergent and convergent evolutions, respectively.
		Note that we do not plot the cases of Runs~2 and 3, because they occupy the same location of Run~1 and indicate the same feature of the evolution as that of Run~1.
		\label{fig:summary_runs}
		}
	\end{center}
\end{figure}
We summarize the features of the orbital evolutions given by our hydrodynamic simulations in Figure~\ref{fig:summary_runs}.
In the figure, we plot a circle for the run in which the evolution of the planet pair is convergent at the end of the simulation.
A cross in the figure indicates the divergent evolution case.
The color in the figure indicates the ratio of $\tmigin$ to $\tmigout$ given by Equation~(\ref{eq:tmig}), which is the same as in Figure~\ref{fig:tmig_ratio_a1e-3_h5e-2}, but the disk parameters are different in the middle and bottom panels.
As can be seen in the figure, all circles are located within the regions where $\tmigin/\tmigout > 1$ (the red filled region).
Most of the crosses are placed within the region where $\tmigin/\tmigout < 1$, except the case of Run~13, for which the migration timescale ratio is close to unity.
Hence, the results of our hydrodynamic simulations agree well with the prediction described in Section~\ref{subsec:prediction_paperI}.

In the case of Run~13, the value of $\tmigin/\tmigout$ is slightly larger than unity (1.05); nonetheless, the evolution of the planet pair is divergent, which is inconsistent with the prediction of Section~\ref{subsec:prediction_paperI}.
But, it confirms our statement given at the end of Section~\ref{subsec:prediction_paperI}, that because of the limited accuracy (within a factor 2--3) of Equation~(\ref{eq:tmig}) in predicting the migration timescales, the value of $\tmigin/\tmigout$ estimated from Equation~(\ref{eq:tmig}) is not accurate when $\tmigin \sim \tmigout$.
In this sense, our prediction is not also decisive in the case of Run~10 because $\tmigin/\tmigout = 1.03$ in this case, though it looks consistent with our prediction.

\subsection{Comparison with the results of BP13} \label{subsec:comparison_BP13}
By carrying out two-dimensional hydrodynamic simulations similar to ours, \citetalias{Baruteau_Papaloizou2013} have shown several examples of the planetary evolution in the disk, in which the initial convergent migration changes at a certain point into the divergent one, and discuss the possible reason for that. 
In the previous sections, we have found out that the prediction described in Section~\ref{subsec:prediction_paperI} agrees reasonably well with the results of our hydrodynamic simulations.
In this subsection, we compare our results to those obtained by \citetalias{Baruteau_Papaloizou2013}.

\citetalias{Baruteau_Papaloizou2013} have presented three typical outcomes of their hydrodynamic simulations in Figure~2 of that paper.
Those results were achieved by adopting $h/R=0.05$ and $\nu=1.1\times 10^{-5}$ which are constant in the whole computational domain.
The value $\nu=1.1\times 10^{-5}$ corresponds to $\alpha=4.5\times 10^{-3}$ at $R=R_0$.
The mass of the outer planet is set to be $\mpout/\mstar=4\times 10^{-4}$.
When $\mpin/\mstar=4.4\times 10^{-4}$ and $\Sigma_0 = 3\times 10^{-4}$ (upper panel of Figure~2 of that paper), the evolution of the planet pair is convergent at the early stage of the simulations.
However, after several hundred orbits, its evolution changes to be divergent, and the period ratio increases with time until the end of the simulations, similarly to the case of Run~1 of our simulations.
Now, we can check what the outcome of these simulations should be according to our phenomenological criterion. 
The ratio of the migration timescales $\tmigin/\tmigout$ evaluated from Equation~(\ref{eq:tmig}) ($\rpin=0.6 R_0$ and $\rpout/\rpin=1.58$ are adopted) is equal to $0.89$.
This is less than unity, so our prediction is consistent with the results of their hydrodynamic simulations.
When $\mpin/\mstar=2.2\times 10^{-4}$ and $\Sigma_0=8\times 10^{-4}$, Equation~(\ref{eq:tmig}) gives $\tmigin/\tmigout=0.80$ ($\rpin=0.6R_0$ and $\rpout/\rpin=1.31$, which corresponds to the 3:2 mean-motion resonance), and hence also in this case, divergent evolution is predicted.
As shown in the lower panel of Figure~2 of \citetalias{Baruteau_Papaloizou2013}, the evolution becomes divergent at the end of the simulations, which is also consistent with the prediction made from Equation~(\ref{eq:tmig}).
When $\mpin=6.6\times 10^{-4}$ and $\Sigma_0=6\times 10^{-4}$ (the middle panel of Figure~2 of \citetalias{Baruteau_Papaloizou2013}), the value of $\tmigin/\tmigout$ obtained from Equation~(\ref{eq:tmig}) is $1.2$ and in the simulations, the period ratio finally reaches $1.67$, which is close to the 5:3 mean-motion resonance.
Also in this case, the prediction from Equation~(\ref{eq:tmig}) and the result of their simulations are consistent with each other.

In Figure~7 of \citetalias{Baruteau_Papaloizou2013}, they have also shown another two examples.
In that figure, the results for lower mass planets ($\mpin/\mstar=4.5\times 10^{-5}$, $\mpout/\mstar=3.9\times 10^{-5}$), smaller aspect ratio and viscosity 
($h/R=0.023$, $\alpha=2.3\times 10^{-3}$) are illustrated.
When the surface density is large ($\Sigma_0=8\times 10^{-5}$), the evolution of the planets finally becomes divergent (the upper panel), while when the surface density is small ($\Sigma_0=3\times 10^{-5}$) the planet pair evolves convergently and is captured into 2:1 mean-motion resonance.
With $\mpin/\mstar=4.5\times 10^{-5}, \mpout/\mstar=3.9\times 10^{-5}$, $h/R=0.023$ and $\alpha=2.3\times 10^{-3}$, the ratio of the predicted migration timescale is smaller than unity ($\tmigin/\tmigout \sim 0.95$).
Hence, the evolution of the planet pair is predicted to be divergent, regardless of the value of $\Sigma_0$.
This prediction seems to be inconsistent with the result in the case of the bottom panel of Figure~7 in \citetalias{Baruteau_Papaloizou2013} (the case of small surface density).
This discrepancy may come from the inaccuracy of our formula for the migration timescale ratios close to unity.
Alternatively, in a less massive disk, the speed of the orbital divergence is small because it is proportional to the migration speed of the planets.
If a longer calculation is done, the evolution may change to be divergent.

Our prediction described in Section~\ref{subsec:prediction_paperI}, except for those cases where the migration times of both planets are comparable to each other, agrees well with the results of \citetalias{Baruteau_Papaloizou2013}, in the same way as with the results of our simulations.

\subsection{A simple criterion for the divergent evolution} \label{subsec:simple_criterion_div_evo}
In summary, the results of the hydrodynamic simulations (both ours and those of \citetalias{Baruteau_Papaloizou2013}) can be easily predicted using our arguments presented in Section~\ref{subsec:prediction_paperI}.
It means that we can have a rough idea of how the planet pair evolves, by comparing their migration timescales calculated from Equation~(\ref{eq:tmig}).
That is, when the migration timescale of the inner planet ($\tmigin$) is much shorter than that of the outer planet ($\tmigout$; both evaluated from Equation~(\ref{eq:tmig})), the evolution of the planet pair is divergent and the period ratio of the planet pair increases with time.
On the other hand, when $\tmigin \gg \tmigout$, the evolution of the planet pair is convergent and the period ratio of the planet pair decreases with time.
This prediction can explain the results of our hydrodynamic simulations, except for the case when $\tmigin \sim \tmigout$.
Hence, we can conclude that if $\tmigin \ll \tmigout$, the evolution of the planet pair is divergent and the planet pair cannot be captured into any mean-motion resonance.
On the other hand, when the planet pair is captured within one of the mean-motion resonances, it is required that $\tmigin \gg \tmigout$.

It should be noted that in this paper, we focus on the cases in which planets form (partial) gaps and the cases in which the mass ratio of the pair does not significantly deviate from unity, i.e., $0.1\lesssim \mpin/\mpout \lesssim 10$ as can be seen in Table~\ref{tab:models}.
In our parameter range, the transition from the convergent to the divergent evolutions can be explained by the change in the migration speed due to the gap formation, as summarized above.
On the other hand, when the inner planet is much smaller or larger than the outer one (for instance, the pair of Jupiter and Earth), the wave lunched by the larger planet may affect the orbital evolution of the smaller one, as shown by \cite{Podlewska-Gaca_Papaloizou_Szuszkiewicz2012}.
Further investigation is required for that parameter range.

\section{Three-body simulations} \label{sec:3body_sims}
\subsection{Numerical method}
As discussed in the previous section, the divergent or convergent character of the planet pair evolution may be predicted on the basis of the ratio of the relevant migration times evaluated from Equation~(\ref{eq:tmig}), with the exception of those cases when $\tmigin \sim \tmigout$.
This gives us the opportunity to investigate the formation of the mean-motion resonances in a broad range of the planet masses, without carrying out the time-consuming hydrodynamic simulations, by incorporating our formula for the migration timescale into the three-body simulations. 
Accordingly, we carry out three-body simulations with our formula for the migration timescale implemented into the code {\sc \tt REBOUND} \citep{Rein_Liu2012} with the {\sc \tt IAS15} integrator \citep{Rein_Spiegel2015}.
We introduce the dissipative forces into the equations of motion which mimic the disk--planet interactions \citep{Lee_Peale2002}.
We use $\tmig$ calculated from Equation~(\ref{eq:tmig_timevar}) which includes the time variation of the migration timescale given by Equation~(\ref{eq:tmig}).
When the planet is small enough, the eccentricity-damping timescale depends on the disk aspect ratio and the migration timescale, namely $\tau_e=c (h/R)^2 \tmig$ with $c=1.28$ \citep[e.g.,][]{Tanaka_Ward2004}.
On the other hand, the time variation of the eccentricity of the gap-opening planet must be different from that of the small planets 
following the linear theory \citep[e.g.,][]{Goldreich_Sari2003,Duffell_Chiang2015}.
Unfortunately, no empirical formula that can be incorporated into the three-body simulations is available.
Due to this fact, as a simple treatment, we adopt the same expression of $\tau_e$ for both the small and the gap-opening planets.
In Appendix~\ref{sec:dependence_tdamp_e}, we present the simulations with larger and smaller values of $c$ and discuss how the evolution of the planet pair is affected by the choice of this parameter.

The initial positions of the inner and outer planets are set to be $\rpin/R_0=1$ and $\rpout/R_0=1.7$.
The masses of the planets and the central star do not change in time.
The initial eccentricities of the planets are set to be $0$.
The distribution of the surface density is $\Sigma_0 (R/R_0)^{-1/2}$ and the disk aspect ratio and the value of $\alpha$ are constant throughout the disk, as in the hydrodynamic simulations presented in Section~\ref{sec:hydro}.

\subsection{Results of three-body simulations}
\subsubsection{A typical evolution of a planet pair} \label{subsubsec:typical_3body_evolutions}
First, we show some typical outcomes of our three-body simulations and compare them with the results of hydrodynamic simulations with the same parameters.
In our three-body simulations, we obtain the convergent evolution and the divergent evolution when $\tmigin \ll \tmigout$ and $\tmigin \gg \tmigout$, 
respectively, exactly the same as in the hydrodynamic simulations.

\begin{figure*}
	\begin{center}
		\resizebox{0.98\textwidth}{!}{\includegraphics{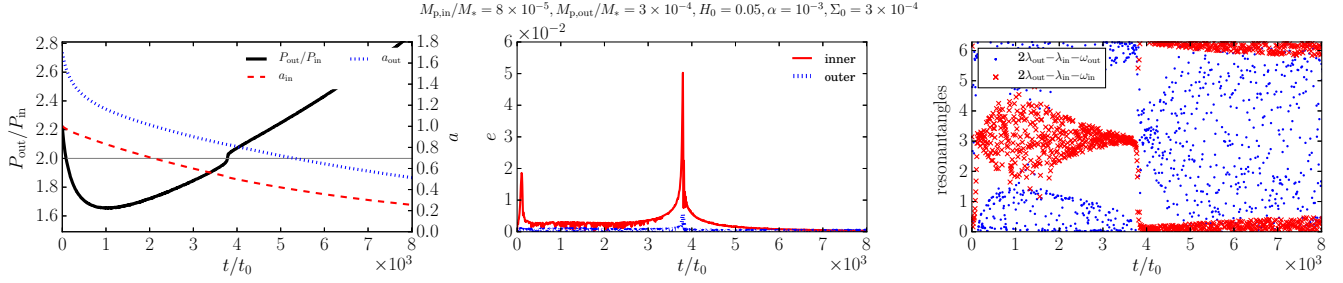}}
		\caption{
		The results of our three-body simulations when $h/R=0.05$, $\alpha=10^{-3}$ and $\Sigma_0=3\times 10^{-4}$.
		The masses of the inner and outer planets are $\mpin/\mstar=8\times 10^{-5}, \mpout/\mstar=3\times 10^{-4}$.
		The gray horizontal line denotes $\pout/\pin=2$.
		The result given by our hydrodynamic simulation with the same parameters are displayed in Figure~\ref{fig:qin8e-5_qout3e-4_h5e-2_a1e-3_S3e-4}.
		\label{fig:nbody_typicals_run1}
		}
	\end{center}
\end{figure*}
Figure~\ref{fig:nbody_typicals_run1} shows the results of our three-body simulations with the same parameters (i.e., planet masses and the values of $h/R$ and $\alpha$) as those in Run~1 (the result given by our hydrodynamic simulation is shown in Figure~\ref{fig:qin8e-5_qout3e-4_h5e-2_a1e-3_S3e-4}).
In both the three-body and hydrodynamic simulations, the period ratio of the planet pair decreases with time until $t\sim 1000\ t_0$.
After that time, the period ratio increases with time and the evolution of the planet pair becomes divergent.
In this sense, the three-body simulation reproduce well the feature of the evolution given by the hydrodynamic simulation, though the details of the evolution are a bit different.
This difference might have originated from the fact that in the three-body simulation, due to our approximation, the migration velocities of the inner and outer planets are slightly slower than those given by the hydrodynamic simulations.
However, what is even more important, the relative migration rate is faster and that is why the planets passed through 2:1 mean-motion resonance as opposed to the hydrodynamic results.

\begin{figure*}
	\begin{center}
 		\resizebox{0.98\textwidth}{!}{\includegraphics{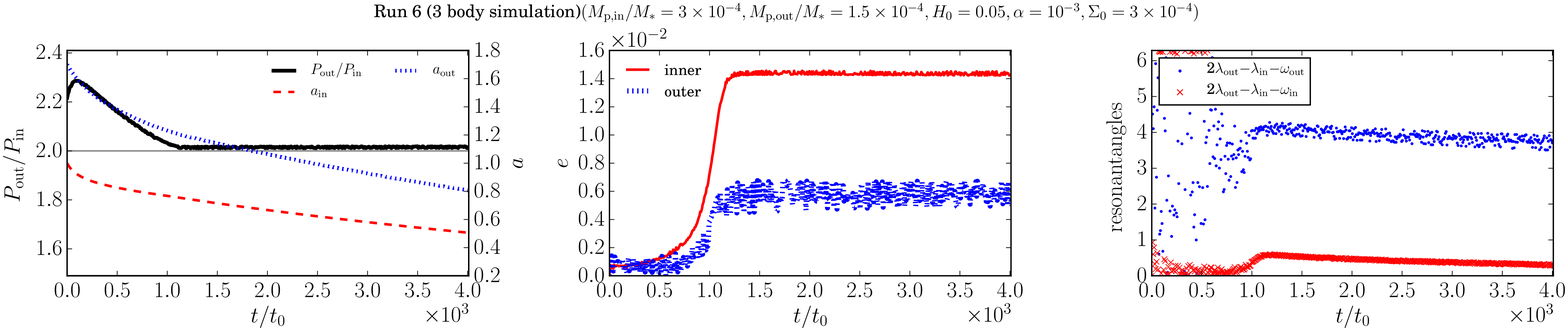}}
 		\resizebox{0.98\textwidth}{!}{\includegraphics{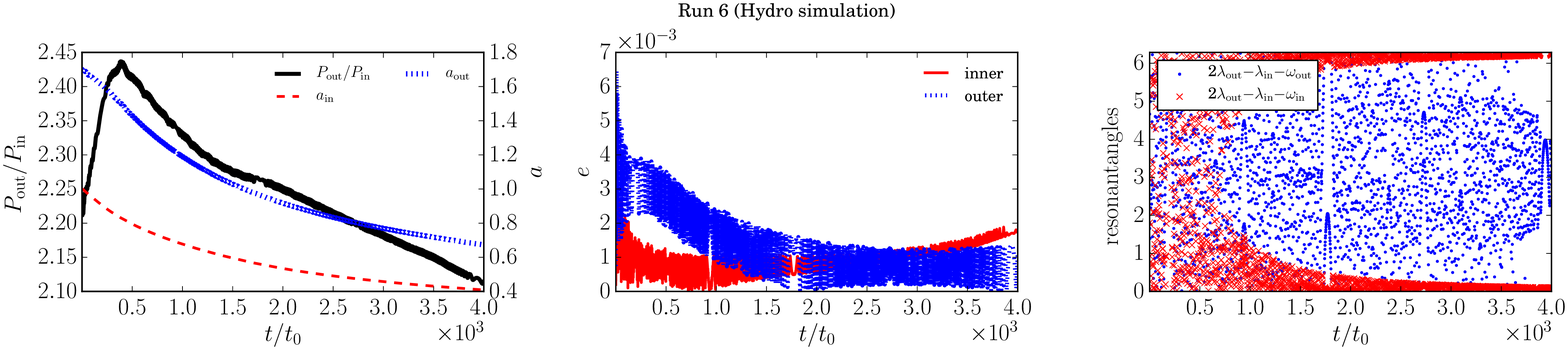}}
		\caption{
		The results of our three-body simulations (upper panels) and hydrodynamic simulations (lower panels) in the case of Run~6.
		The left, middle and right panels indicate the time variations of the semi-major axes and period ratio, the eccentricity, and the resonant angles for the 2:1 mean-motion resonance, respectively.
		\label{fig:nbody_typicals_run6}
		}
	\end{center}
\end{figure*}
Figure~\ref{fig:nbody_typicals_run6} compares the results of three-body simulations and hydrodynamic simulations, in the case of Run~6.
In the case of the three-body simulation, the period ratio decreases in time until $t\sim 1000\ t_0$ and then the planet pair is trapped in 2:1 mean-motion resonance.
On the other hand, in the case of the hydrodynamic simulation, the period ratio also decreases in time, but unfortunately, at the time when our hydrodynamic simulations ended (at $t=4000\ t_0$), the planet pair still has not arrived at the location of the 2:1 mean-motion resonance, so more detailed comparison is not possible.
However, there is a significant indication that the planets are approaching the 2:1 mean-motion resonance, judging from the behavior of the resonance angle.
Therefore, it is likely that the hydrodynamic and three-body simulations give the convergent evolution and qualitatively agree with each other.

As shown above, our three-body simulations can reproduce qualitatively the cases of the divergent evolution and convergent evolution given by our hydrodynamic simulations.
Keeping in mind the inaccuracy of Equation~(\ref{eq:tmig_timevar}) and the simple treatment of the eccentricity damping, it is expected that the details of the evolutions of the semi-major axes and the eccentricities of the planet pair in the results of the three-body simulations and the hydrodynamic simulations may differ. 
However, both in the three-body simulations and the hydrodynamic simulations, the evolution of the planet pair becomes divergent when $\tmigin \gg \tmigout$, and it becomes convergent when $\tmigin \ll \tmigout$.

\subsubsection{A parameter survey in a wide range of masses of the planet pair} \label{subsubsec:3body_param_servey}
Varying the masses of the inner and outer planets in a wide range with the fixed values of the disk parameters $h/R$ and $\alpha$, we examine the period ratios of the planet pairs at the end of the simulations.

In the following, we calculate the radial migrations of the planets until the time will reaches the value $\tmigin$ evaluated at the initial position of the inner planet.
The semi-major axis of the inner planet can be given by $R_0 \exp(-t/\tmigin)$ if the effect of the outer planet is negligible (or the inner planet is isolated). 
At $t=\tmigin$ (at the end of the simulation), hence, the position of the inner planet is expected to be $\rpin \simeq 0.36R_0$.
It is sufficiently long in order to examine the characteristics of the evolutions of the planet pair, which is convergent or divergent.
When the inner planet is strongly pushed by the outer planet, it reaches the inner part of the disk, namely $R=0.36R_0$, before $t=\tmigin$.
In this case, we terminate the simulations when $\rpin$ becomes smaller than $0.3R_0$.

\begin{figure}
	\begin{center}
		\resizebox{0.48\textwidth}{!}{\includegraphics{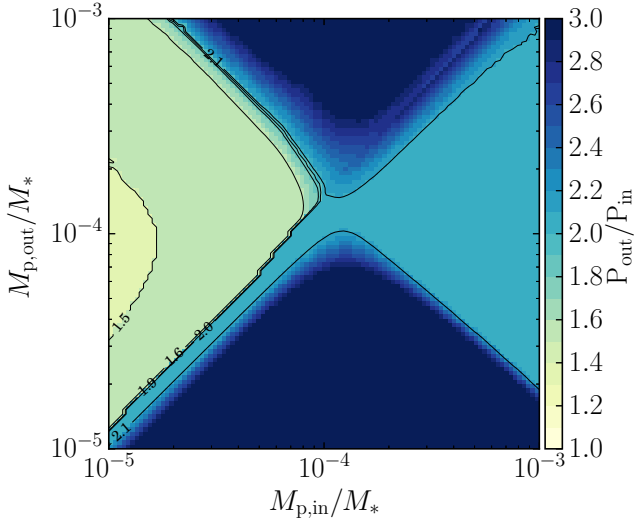}}
		\caption{
		The values of $\pout/\pin$ at the end of the three-body simulations when $h/R=0.05$, $\alpha=10^{-3}$ and $\Sigma_0=3\times 10^{-4}$.
		\label{fig:period_ratio_h0.05_a1e-3}
		}
	\end{center}
\end{figure}
Figure~\ref{fig:period_ratio_h0.05_a1e-3} shows the period ratios of the planet pair at the end of the three-body simulations when $h/R=0.05$ and $\alpha=10^{-3}$.
In the calculations presented in this figure, we adopt $\Sigma_0=3\times 10^{-4}$.
When $\mstar=1M_{\odot}$ and $R_0=10 \au$, this value of $\Sigma_0$ corresponds to $27\mbox{ g/cm}^2$, which is about half the surface density of 
the minimum solar nebula disk at $10\au$.
Comparing Figure~\ref{fig:period_ratio_h0.05_a1e-3} and Figure~\ref{fig:tmig_ratio_a1e-3_h5e-2}, we find that the results of our three-body simulations agree with the prediction in Section~\ref{subsec:prediction_paperI}.
That is, the evolution of the planet pair is convergent and the planet pair is locked into the resonance if $\tmigin/\tmigout<1$.
When $\mpin<\mptrans$, the period ratio of the planet pair is likely to be $1.6$ -- $1.5$.
Only the planet pair whose masses are similar to each other has the period ratio close to two.
In the case of $\mpin>\mptrans$, on the other hand, most of the planet pairs are locked into the 2:1 mean-motion resonance if the evolution of the planet pair is convergent.
The capture into the mean-motion resonance is discussed in Section~\ref{subsec:resonant_capture}.

\begin{figure}
	\begin{center}
		\resizebox{0.48\textwidth}{!}{\includegraphics{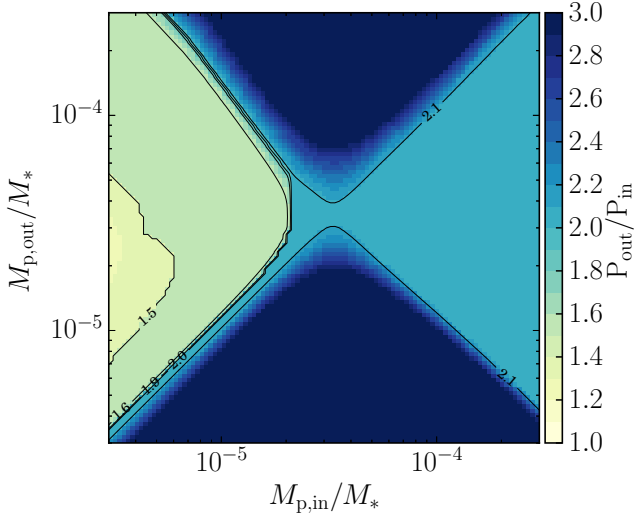}}
		\caption{
		The same as in Figure~\ref{fig:period_ratio_h0.05_a1e-3}, but in the case with $h/R=0.03$, $\alpha=10^{-3}$ and $\Sigma_0=1\times 10^{-4}$.
		\label{fig:period_ratio_h0.03_a1e-3}
		}
	\end{center}
\end{figure}
As discussed in Section~\ref{subsec:prediction_paperI}, in the inner region of the disk with the small disk aspect ratio, the planetary migration slows down due to the gap formation, even if the planet mass is in the range of the super-Earth.
In Figure~\ref{fig:period_ratio_h0.03_a1e-3}, we show the period ratios given by the three-body simulations adopting Equation~(\ref{eq:tmig_timevar}) when $h/R=0.03$, $\alpha=10^{-3}$ and the planet masses are in the mass range ($3\times 10^{-6} < \mpl/\mstar < 3\times 10^{-4}$) including the range of the super-Earth.
In this case, $\Sigma_0=1\times 10^{-4}$ or $889 \mbox{ g/cm}^2$ (when $\mstar=1M_{\odot}$ and $R_0=1\au$) is adopted, which is also about half 
the surface density of the minimum solar nebula disk at $1\au$.
As in the previous calculations presented in Figure~\ref{fig:period_ratio_h0.05_a1e-3}, also here when the mass of the inner planet is large enough ($\mpin > \mptrans$), most of the planet pairs are locked into the 2:1 mean-motion resonance.
On the other hand, when the mass of the inner planet is smaller than $\mptrans$, the planet pair can easily pass through the 2:1 mean-motion resonance.

\section{Discussion} \label{sec:discussion}
\subsection{Capture into the mean-motion resonance} \label{subsec:resonant_capture}
When the evolution of the planet pair is convergent, the planet pair can be captured into mean-motion resonance.
\cite{Ogihara_Kobayashi2013} investigated the condition for the capture into first order mean-motion resonance.
They found that when the relative migration timescale is longer than a critical timescale, the planet pair is captured into that resonance.
This critical timescale is given by 
\begin{align}
t_{\rm crit} &= C \bracketfunc{\mpin}{\mstar}{-4/3} \omegain^{-1},
\label{eq:tcrit}
\end{align}
where $C=0.27$ for the 2:1 mean-motion resonance, and $C=0.054$ for the 3:2 mean-motion resonance, in the case of $\mpout/\mpin \sim 1$.
When the value of $\mpin/\mpout$ is much smaller or larger than unity, the coefficient of $C$ is larger.
Note that \cite{Ogihara_Kobayashi2013} investigated the cases of $\mpout/\mpin<1$.
However, we checked that their results can be extended to the case of $\mpout/\mpin>1$ (Appendix~\ref{sec:tcrit}). 
We confirmed that in most cases shown in Figures~\ref{fig:period_ratio_h0.05_a1e-3} and \ref{fig:period_ratio_h0.03_a1e-3}, when the planet pair is captured into the 2:1 mean-motion resonance, the relative migration timescale is longer than the critical timescale.
When the planet pair is captured into the 3:2 mean-motion resonance, the relative migration timescale is shorter than the critical timescale for the 2:1 mean-motion resonance, but it is longer than the critical timescale for 3:2 mean-motion resonance. 
\footnote{
We compute the relative migration timescale by $(\rpin+\rpout)/(2 v_{\rm rel})$, where $v_{\rm rel} = \rpout/\tmigout - \rpin/\tmigin$.
Strictly speaking, the values of $\rpin$ and $\rpout$ at the time when the planet pair passes through the resonant location are required to compute the above relative migration timescale.
For simplicity, we adopt $\rpin = 1.0$ and $\rpout=1.6$, instead of exact values.
This simplification does not change the relative migration timescale given by three-body simulations, because $\tmigin$ and $\tmigout$ are independent of $R$ with our disk model (i.e., $s=0.5$ and $f=0$).
It just slightly affects the critical timescale in $\omegain$.
}
Hence, we conclude that the results of our three-body simulations are consistent with the results of \cite{Ogihara_Kobayashi2013}.

The feature of the resonant capture shown in Figures~\ref{fig:period_ratio_h0.05_a1e-3} and \ref{fig:period_ratio_h0.03_a1e-3} can be explained by the dependence of the critical timescale on the mass of the inner planet.
According to Equation~(\ref{eq:tcrit}), the critical timescale becomes shorter as the mass of the inner planet increases.
When $\mpin < \mptrans$, the planet pair can be captured into 3:2 mean-motion resonance rather than  2:1 resonance, because the critical timescale for the 2:1 mean-motion resonance is long in view of the fact that $\mpin$ is small.
As $\mpin$ increases, the critical timescale becomes shorter.
Hence, when $\mpin>\mptrans$, the planet pair has a bigger chance of being trapped in 2:1 resonance than when $\mpin<\mptrans$ (see also Appendix~\ref{sec:tcrit}). 
Because of it, in the region where $\mpin<\mptrans$, the planet pair undergoing the convergent migration is likely to be captured into 3:2 resonance or other commensurability with the higher value of integers (as for example 4:3), while when $\mpin>\mptrans$, it is expected that it will be captured into 2:1 mean-motion resonance.

In our hydrodynamic simulations, the mean-motion resonance into which the planets are captured is the same as that obtained in three-body simulations, except for two cases, namely, Run~7 and Run~12.
For instance, in the case of Run~7, the planet pair is captured into the 3:2 mean-motion resonance as can be seen in Figure~\ref{fig:orbitevo_runs4-15}, while in the three-body simulations the planet pair ends up into 2:1 mean-motion resonance as can be seen from Figure~\ref{fig:period_ratio_h0.05_a1e-3}.
\begin{figure}
	\begin{center}
		\resizebox{0.48\textwidth}{!}{\includegraphics{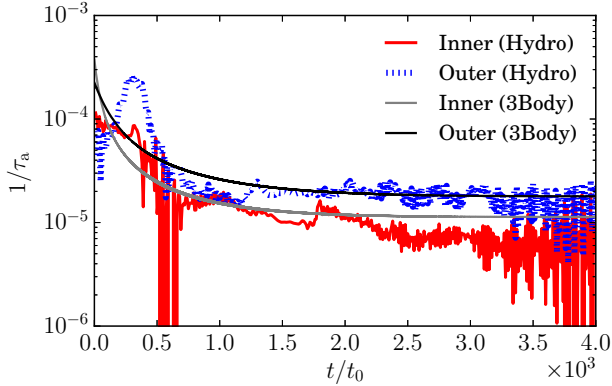}}
		\caption{
		Time variations of the migration timescales in the case of Run~7.
		The red solid line and blue dashed line denote the results obtained in hydrodynamic simulations for the inner and outer planets, respectively, and the gray and black thin lines represent outcomes of three-body simulations for the inner and outer planets, respectively.
		\label{fig:tmig_Run7}
		}
	\end{center}
\end{figure}
Figure~\ref{fig:tmig_Run7} shows the time variations of the migration timescales for the planet pairs, resulting from hydrodynamic simulations and three-body simulations, in the case of Run~7.
The overall evolution of the migration timescale in the three-body and hydrodynamic simulations agrees reasonably well with each other.
The same can be concluded for the relative velocity of the planet pair, except at the early phase, $400<t/t_0<500$, when both the migration of the outer planet and the planets' relative velocity is fast.  
During this short phase, the planet pair passes through 2:1 mean-motion resonance.
The difference in the type of resonance in which planets end up in our calculations is caused by the described dissimilarities in the early evolution.

When the mass accretion onto the planet is considered, the resonant capture can be affected by the change of the planet mass.
Once the mass of the planet reaches the critical core mass (typically $\sim 10 M_{\oplus}$ \citep{Mizuno1980,Kanagawa_Fujimoto2013}), the planetary mass increases quickly due to the onset of the runaway gas accretion and the planet becomes a giant \citep[e.g.,][]{Ida_Tanaka_Johansen_Kanagawa_Tanigawa2018,Tanaka_Murase_Tanigawa2019}.
However, the critical core mass can vary up to $\sim 50 M_{\oplus}$, depending on the accretion rate of the planetesimals and opacity of the atmosphere \citep[e,g,][]{Rafikov2006}.
Moreover, recent 3D hydrodynamic simulations \citep{Lambrechts_Lega_Nelson_Crida_Morbidelli2019} have shown that the quasi-static contraction during the runaway gas accretion can be much slower than that expected by 1D models \citep[e.g.,][]{Pollack_etal1996,Hubickyj_Bodenheimer_Lissauer2005}, when the mass of the planet is smaller than Saturn.
Indeed, a number of exoplanets with an intermediate mass, namely, $10M_{\oplus}$ to $100 M_{\oplus}$, have been observed by, e.g., the Kepler telescope.
Such an intermediate mass implies an inefficient mass growth.
The mechanism of the gas accretion onto the planet is not fully understood yet as described above.
The mass growth mechanism will be considered in future work.

\subsection{Implication for the formation of planetary systems} \label{subsec:implications}
As discussed in Section~\ref{subsec:simple_criterion_div_evo}, we conclude that the transition from the convergent evolution to the divergent evolution is caused by the slowdown of the migration speed of the outer planet due to the gap opening, rather than the planet--wake interaction and other hydrodynamic effects.
Hence, after the migration reaches the stationary speed, the planet pair cannot escape from the resonance by this process, if the planet mass and disk parameters are not changed.
However, the transition from convergent to divergent evolution can occur when the planet mass increases during the migration.
For instance, in the case of $\mpin,\mpout>\mptrans$, the divergent evolution can occur when the growth timescale of the inner planet is longer than that of the outer planet.
If the formation of one planet took place much earlier (later) than that of another planet, the transition from the convergent to the divergent evolutions (divergent to  convergent evolutions) may occur.
In the inner region of the disk, moreover, as the value of $\mptrans$ becomes small because the disk aspect ratio is small, the migration speed becomes slower as the planet migrates inward.
As a result, a transition from the divergent to the convergent evolutions may happen because the inner planet slows down due to the gap opening as compared to the migration of the outer planet.
In this case, the planet pair is captured into the mean-motion resonance, even when the evolution of the planet pair is divergent in the outer region.
Depending on the distribution of the disk parameters, the transition from the convergent to the divergent evolution may be possible by the same mechanism.
The condition of the transition of the migration feature can be obtained from our migration model given by Equation~(\ref{eq:tmig}).
This effect can affect the formation of the planet pair in the mean-motion resonance including the close-in planets observed by the Kepler telescope.

Our migration model described in Section~\ref{subsec:prediction_paperI} provides the condition for divergent evolution, which is consistent with the results of hydrodynamic simulations as discussed in Section~\ref{subsec:simple_criterion_div_evo}.
Our model can also provide the condition for the convergent evolution, during which the planet pair can be captured into the mean-motion resonance.
As discussed in Section~\ref{subsec:resonant_capture}, by combining the critical timescale provided by \cite{Ogihara_Kobayashi2013}, we may be able to predict in which resonance the planets can be captured.
Moreover, as shown by \cite{Izidoro_etal2017} and \cite{Ogihara_Kokubo_Suzuki_Morbidelli2018}, the planet pair captured into the mean-motion resonance can be unlocked by the onset of dynamical instability after the dispersal of the gaseous disk.
For the onset of the dynamical instability, the separation between the planet pair, which is a consequence of the radial migration, is essential \citep[e.g.,][]{Chambers_Wetherill_Boss1996,Marzari_Weidenschilling2002,Wu_Zhang_Zhou_Steffen2019}.
When the transition from convergent to divergent evolution occurs, the stability of the system would be significantly changed.
In this sense, our work would be helpful to explain the observed distribution of the period ratio.
However, our simulation does not take into account any processes of the dissipation in the gaseous disk.
We also consider only systems with two planets, whereas in general, exoplanetary systems contain more planets.
Because the stability of the system depends on the number of the planets \citep[e.g.,][]{Chambers_Wetherill_Boss1996,Simbulan_Tamayo_Petrovich_Rein_Murray2017,Matsumoto_Kokubo2017}, it is necessary to consider the cases with three or more planets to investigate the effects on stability by incorporating our model into population synthesis calculations, such as \cite{Mordasini_Alibert_Benz_Klahr_henning2012} and \cite{Ida_Lin_Nagasawa2013}, which will be done in future works.

We should note that nonisothermal effects are very important to understand the planetary migration, especially the type~I regime as shown by e.g., \cite{Paardekooper_Baruteau_Crida_Kley2010,Bitsch_Johansen_Lambrechts_Morbidelli2015}.
When the gaps created by each planet in the pair merge together and form a common gap, the migration speed of the planets could deviate from that of the single planet predicted by our model (Equation~(\ref{eq:tmig})).
As discussed by \cite{Tanigawa_Tanaka2016}, the gas accretion onto the planet may change the entire structure of the disk, which should be considered.
Moreover, when the orbital inclination is highly excited by the resonant capture \citep[e.g.,][]{Thommes_Lissauer2003,Teyssandier_Terquem2014}, the migration timescale of the gap-opening planet can differ from that given by Equation~(\ref{eq:tmig}).
The migration time can be shorter because the gap is shallower for the planet with a larger inclination \citep[e.g.,][]{Bitsch_Crida__Libert_Lega2013,Chametla_Sanchez_Masset_Hidalgo2017,Zhu2019}, while it can be longer because the disk--planet interaction itself is weaker as the inclination increases \citep[e.g.,][]{Rein2012b,Arzamasskiy_Zhu_Stone2018}.
The migration of the gap-opening planet with a finite inclination would be determined by the balance between the two effects above, whereas it can be given by Equation~(\ref{eq:tmig}) when the inclination is not that large.
In this paper, we focus on the evolution of the planet pair whose mass ratio does not differ much from unity.
When the mass ratio is much larger/smaller than unity, for instance, the Jupiter--Earth system, the planet--wake interaction may be more effective than that in our cases, as shown by \cite{Podlewska-Gaca_Papaloizou_Szuszkiewicz2012}.
The effect of gas self-gravity can modify the migration velocity \citep[e.g.,][]{Baruteau_Meru_Paardekooper2011}, and it may change the commensurability of a resonance in which the planets are locked \citep{Ataiee_Kley2020}.
However, general trends that we found in this paper, i.e., the transition of the convergent to divergent evolutions, the condition of the resonant capture, qualitatively would not change, though the commensurability of the resonance shown in Figures~\ref{fig:period_ratio_h0.05_a1e-3} and \ref{fig:period_ratio_h0.03_a1e-3} might be affected.
Further investigation is required for the full understanding of the effects of the above processes on the occurrence of the resonances in planetary systems.

\subsection{Comparison with observations} \label{subsec:comp_obs}
\begin{figure}
	\begin{center}
		\resizebox{0.48\textwidth}{!}{\includegraphics{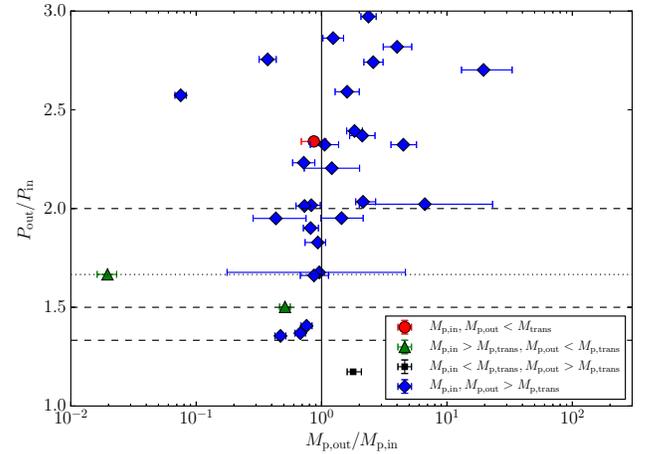}}
		\caption{
		The period and mass ratios for the inner and outer planets in the systems, in which only two planets are observed at present.
		The red circles indicate systems in which the masses of the inner and outer planets are smaller than $\mptrans$. 
		In order to estimate $\mptrans$, we assume $\alpha =10^{-3}$, $h/R =0.03$ (if the orbital radius is smaller than 1AU) and $h/R = 0.03(R/1AU)^{1/4}$ (if the orbital radius is larger than or equal to 1AU).
		The blue diamonds represent the cases in which both the masses of both the inner and outer planets are larger than $\mptrans$.
		The green triangles denote the case in which $\mpin > \mptrans$ and $\mpout < \mptrans$, instead the black squares in which  in which $\mpin < \mptrans$ and $\mpout > \mptrans$.
		The horizontal dashed lines indicate fist-order mean-motion resonances, the 2:1, the 3:2 and the 4:3 commensurabilities from the top, respectively.
		The horizontal dotted lines indicate the location of the 5:3 mean-motion resonance.
		\label{fig:massratio_vs_pratio_comp_with_obs}
		}
	\end{center}
\end{figure}
Using our results, we may check whether we can find in the observed distribution of the period ratios of the two-planet systems any characteristic features that originated during the  early phase of the planetary migration. 
With this scope in mind, in Figure~\ref{fig:massratio_vs_pratio_comp_with_obs}, we plot the relation between the period ratio and the mass ratio in the observed two-planet systems extracted from the NASA Exoplanet Archive \footnote{\url{https://exoplanetarchive.ipac.caltech.edu/}} \citep{Akeson2013}.
We select  those systems around a single star, in which only two planets have been observed until now, in order to exclude additional effects that are important if more planets are present.
For example, in the systems in which there are three or more planets, the architecture could be significantly affected by the orbital instability.
We also excluded planet pairs in which the mass of either planet is larger than $10M_J$.
The selected systems are listed in Table~\ref{tab:2planet_sys} in Appendix \ref{sec:list_twoplanet_sys}.
In drawing Figure~\ref{fig:massratio_vs_pratio_comp_with_obs}, we estimate $\mptrans$ assuming a flaring disk with $h/R=0.03 (R/1{\rm AU})^{1/4}$ for $R>1{\rm AU}$ and a constant disk aspect ratio with $h/R=0.03$ for $R<1{\rm AU}$ considering the disk structure of the inner rim \citep[e.g.,][]{Flock_Fromang_Turner_Benisty2016,Ueda_Okuzumi_Flock2017}.
$\mptrans$ is calculated using the value of $h/R$ at the location of the inner planet. 
The adopted value of the $\alpha$ is $10^{-3}$.
Because most of the planets are larger than $\mptrans$ in the systems shown in Figure~\ref{fig:massratio_vs_pratio_comp_with_obs}, the most of the systems correspond to the case of $\mpin,\mpout > \mptrans$ (blue diamonds, category 4 in Table~\ref{tab:2planet_sys}).

In the case of $\mpin,\mpout > \mptrans$, our results indicate that when $\mpout/\mpin < 1$, the planet pair is likely to be captured in the resonance, whereas the planet pair is unlikely to be captured when $\mpout/\mpin > 1$.
As can be seen in Figure~\ref{fig:massratio_vs_pratio_comp_with_obs},  when $\mpout/\mpin < 1$, all blue diamonds, except for three systems, are captured into the resonance.
Those that are not locked in any commensurability are the following: HD~45184 ($\mpout/\mpin=0.7$), OGLE-2006-BLG-109L ($\mpout/\mpin=0.4$), and rho~CrB ($\mpout/\mpin=0.1$).
For HD~45184, the mass ratio of the outer to the inner planets is close to unity, and our prediction is not very accurate in this range of the mass ratio.
For OGLE-2006-BLG-109L, the period ratio is not precisely known and it can be $1.33 < \pout/\pin < 3.62$; hence, it may be consistent with our prediction.
The planetary system around rho~CrB is composed of a Jupiter-mass planet and a Neptune-size planet.
In such a system, planet--wake interaction may affect the orbital evolution of the planets as shown by \cite{Podlewska-Gaca_Papaloizou_Szuszkiewicz2012}.
Alternatively, it may be formed by trapping the outer planet at the edge of the gap formed by the inner planet \citep{Pierens_Nelson2008}.

In the case of $\mpout/\mpin > 1$, among the blue diamonds in Figure~\ref{fig:massratio_vs_pratio_comp_with_obs}, there are two systems close to 2:1 mean-motion resonance.
This fact is not in conflict with our results, because the period ratio can be around two when $\mpout/\mpin \sim 1$ and $\mpin,\mpout > \mptrans$ (see the upper right region of Figure~\ref{fig:period_ratio_h0.05_a1e-3}).
Alternatively, it may indicate the effect of the common gap formation (Appendix~\ref{sec:common_gap}).
Other systems are distributed above the 2:1 mean-motion resonance, which are consistent with our results.

We have found only a few systems in three other categories defined according to the inner and outer planet mass relation to the $\mptrans$.
There is only one system in the category 1: $\mpin,\mpout < \mptrans$ (the red circle in Figure~\ref{fig:massratio_vs_pratio_comp_with_obs}), two systems belong to the category 2: $\mpin>\mptrans$ and $\mpout<\mptrans$ (the green triangles) and again only one is in the category 3: $\mpin<\mptrans$ and $\mpout>\mptrans$ (the black square).
There is not sufficient statistics in order to make a  decisive conclusion, but these system migration histories are consistent with our predictions, except for KOI-1599 (the green triangle at $\mpout/\mpin=0.5$).
\cite{Panichi_Migaszewski_Gozdziewski2019} have shown that the planetary system of KOI-1599 can be explained by the migration capture when the migration timescale of the inner planet is longer than that of the outer planet.
This condition can be satisfied for different disk parameters from those used in drawing Figure~\ref{fig:massratio_vs_pratio_comp_with_obs}.
Modeling the migration histories of the individual sources can be a natural extension of this work. 

We should note that the classification shown in Figure~\ref{fig:massratio_vs_pratio_comp_with_obs} depends on the aspect ratio and the viscosity.
With smaller viscosity and $h/R$, the mass of the planet can be larger than $\mptrans$, because $\mptrans$ becomes smaller.
In the case presented here, almost all of the systems are classified as the category 4, in which $\mpin,\mpout > \mptrans$ (blue diamonds).
If the viscosity and $h/R$ will be larger, several blue diamonds can be shifted to another category.
However, we confirmed that the general trend as mentioned above does not change if a relatively high viscosity ($\alpha=10^{-2}$) is adopted. 

\section{Summary} \label{sec:summay}
We have investigated the radial migration of the planet pairs in the protoplanetary disks by carrying out the hydrodynamic simulations and the three-body simulations.
Our results are summarized as follows:
\begin{enumerate}
  \item The divergent or convergent character of the radial evolution of the planet pair can be roughly predicted by using the formula of the migration timescale for a single planet embedded in the disk, given by Equation~(\ref{eq:tmig}), as discussed in Section~\ref{subsec:prediction_paperI}. 
  If the ratio of the migration timescales of the inner planet to those of the outer planet ($\tmigin/\tmigout$) is larger than unity, the evolution of the planet pair is expected to be convergent. 
  If $\tmigin/\tmigout<1$, the evolution of the planet pair is expected to be divergent.
  The results of our hydrodynamic simulations shown in Section~\ref{sec:hydro} agree well with the above prediction.
  \item Even when the evolution of the planetary pair is genuinely divergent, which means that at the end of a sufficiently long calculation the planets migrate away from each other, the planet pair can enter the mean-motion resonance before the gap structure reaches steady state. 
	In this case, this planet pair can be temporarily locked into the mean-motion resonance. 
	However, the migration of the outer planet eventually will slow down due to the gap formation. 
	As a result, the planet pair leaves from the resonance position.
	This transition from convergent to divergent evolution can be explained by gap formation as discussed in Section~\ref{sec:hydro}, rather than by planet--wake interaction and other hydrodynamic effects.
  \item We have incorporated our migration model given by Equation~(\ref{eq:tmig_timevar}) into the three-body simulations and obtained the divergent and convergent evolutions of the planet pairs, under conditions similar to those obtained from the hydrodynamic simulations (Section~\ref{sec:3body_sims}).
  \item Our results indicate that after the gaps reach a stationary structure, the planet pair does not escape from the mean-motion resonance.
  However, when the masses of the planet pair increase and the disk parameters (viscosity and disk aspect ratio) change as the planet pair migrates, the escape from the resonance can occur.
  This effect can contribute to the explanation of the distribution of the period ratios of the planet pairs observed by the Kepler.
\end{enumerate}

\acknowledgements
We would like to thank the anonymous referee for his/her constructive and valuable comments which were very helpful in improving the manuscript.
This work was supported by the Polish National Science Centre MAESTRO grant DEC-2012/06/A/ST9/00276.
KDK was also supported by JSPS Core-to-Core Program ``International Network of Planetary Sciences'' and JSPS KAKENHI grant  19K14779.
Numerical computations were carried out on the Cray XC50 at the Center for Computational Astrophysics, National Astronomical Observatory of Japan and the computational cluster of Research Center for the Early Universe.
Three-body simulations in this paper made use of the REBOUND code which can be downloaded freely at \url{http://github.com/hannorein/rebound}.

\software{FARGO \citep{Masset2000}, Matplotlib \citep[\url{http://matplotlib.org}]{Matplotlib}, NumPy \citep[\url{http://www.numpy.org}]{NumPy}}

\appendix

\section{Common gap formation} \label{sec:common_gap}
Here we briefly discuss the effect of a merging gap of a planet pair on the evolution of the period ratio.
When the gaps merge together to form a common gap, the depth and width of the common gap is significantly different from those of the gap formed by a single planet, as shown by \cite{Duffell_Dong2015}.
In this case, the migration speed of the planet would be different from that expected from Equation~(\ref{eq:tmig}).
Moreover, it is also possible that the one planet is trapped at the edge of the gap formed by the other planet 
\citep[e.g.,][]{Pierens_Nelson2008,Podlewska_Szuszkiewicz2009,Cimerman_Kley_Kuiper2018}.

To show the effect of the formation of the common gap, we carry out the simulations, varying the initial position of the outer planet, with the same planet masses and the disk parameters (i.e. $h/R$ and $\alpha$) as those in Run~1.
To avoid the initial convergent evolution, we initially construct the gaps in the disk around the inner and outer planets, using the model of \cite{Kanagawa2017b} (for detail, see Section 4.3 of \citetalias{PaperI}).
\begin{figure}
	\begin{center}
		\resizebox{0.48\textwidth}{!}{\includegraphics{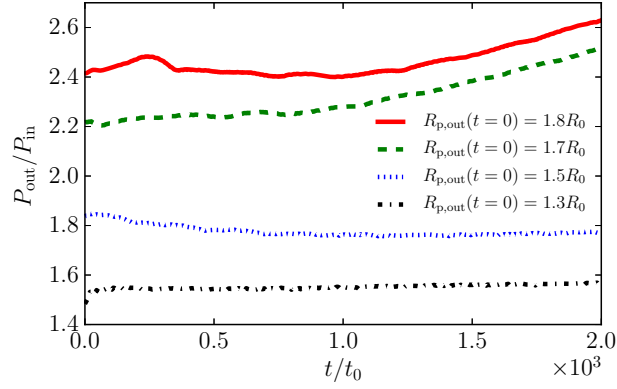}}
		\caption{
		The time variations of the period ratio given by the simulations with the initial gap.
		The initial position of the outer planet ($\rpout(t=0)$) is different in each run: from the top, the cases of $\rpout(t=0)=1.8R_0$, $1.7R_0$, $1.5R_0$, and $1.3R_0$, respectively.
		\label{fig:evo_period_ratio_winitgap}
		}
	\end{center}
\end{figure}
Figure~\ref{fig:evo_period_ratio_winitgap} shows the time variations of the period ratio for the simulations with $\rpout(t=0)=1.8R_0$, $\rpout(t=0)=1.7R_0$, 
$\rpout(t=0)=1.5R_0$, and $\rpout(t=0)=1.3R_0$.
The inner planet is always placed at $R_0$.
As can be seen in the figure, the period ratio increases with time in the cases of $\rpout(t=0)=1.8R_0$ and $\rpout(t=0)=1.7R_0$.
On the other hand, when $\rpout(t=0)=1.5R_0$ and $\rpout(t=0)=1.3R_0$, the period ratios do not significantly change in time.

\begin{figure}
	\begin{center}
		\resizebox{0.48\textwidth}{!}{\includegraphics{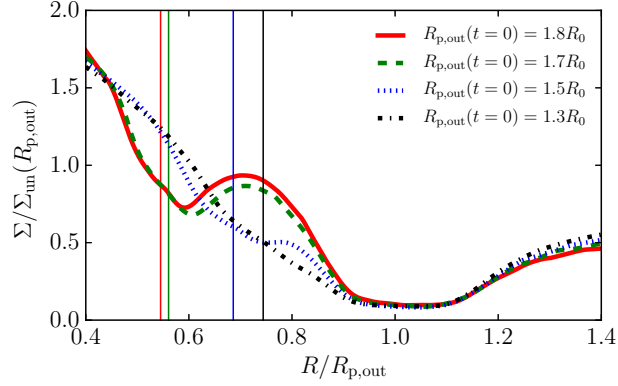}}
		\caption{
		The distribution of azimuthally averaged surface density in the same cases as those presented in Figure~\ref{fig:evo_period_ratio_winitgap}, at $t=1500\ t_0$.
		The vertical thin lines indicate the orbital radius of the inner planet in the cases of $\rpout(t=0)=1.8R_0$, $1.7R_0$, $1.5R_0$, and $1.3R_0$, from the left, respectively.
		\label{fig:avgdens_winitgap}
		}
	\end{center}
\end{figure}
Figure~\ref{fig:avgdens_winitgap} shows the radial distributions of azimuthally averaged surface density at $t=1500\ t_0$ for the cases presented in Figure~\ref{fig:evo_period_ratio_winitgap}.
Since the distance between the inner and outer planets is large enough in the cases of $\rpout(t=0)=1.7R_0$ and $1.8R_0$, the shapes of the gaps are very similar in those two cases.
Instead, in the case of $\rpout(t=0)=1.3R_0$, the inner and outer planets form the common gap.
The case of $\rpout(t=0)=1.5R_0$ is an intermediate case.
When the gaps formed by two planets are not clearly separated, though it is not exactly the value for the mean-motion resonance, the period ratio is not changed from the initial value.
In this case, the inner planet migrates as it is locked into the gap edge.

\section{Dependence on the damping timescale of the eccentricity}
\label{sec:dependence_tdamp_e}
Here we discuss the dependence of the orbital evolution on the damping timescale of the eccentricity.
\begin{figure}
	\begin{center}
		\resizebox{0.48\textwidth}{!}{\includegraphics{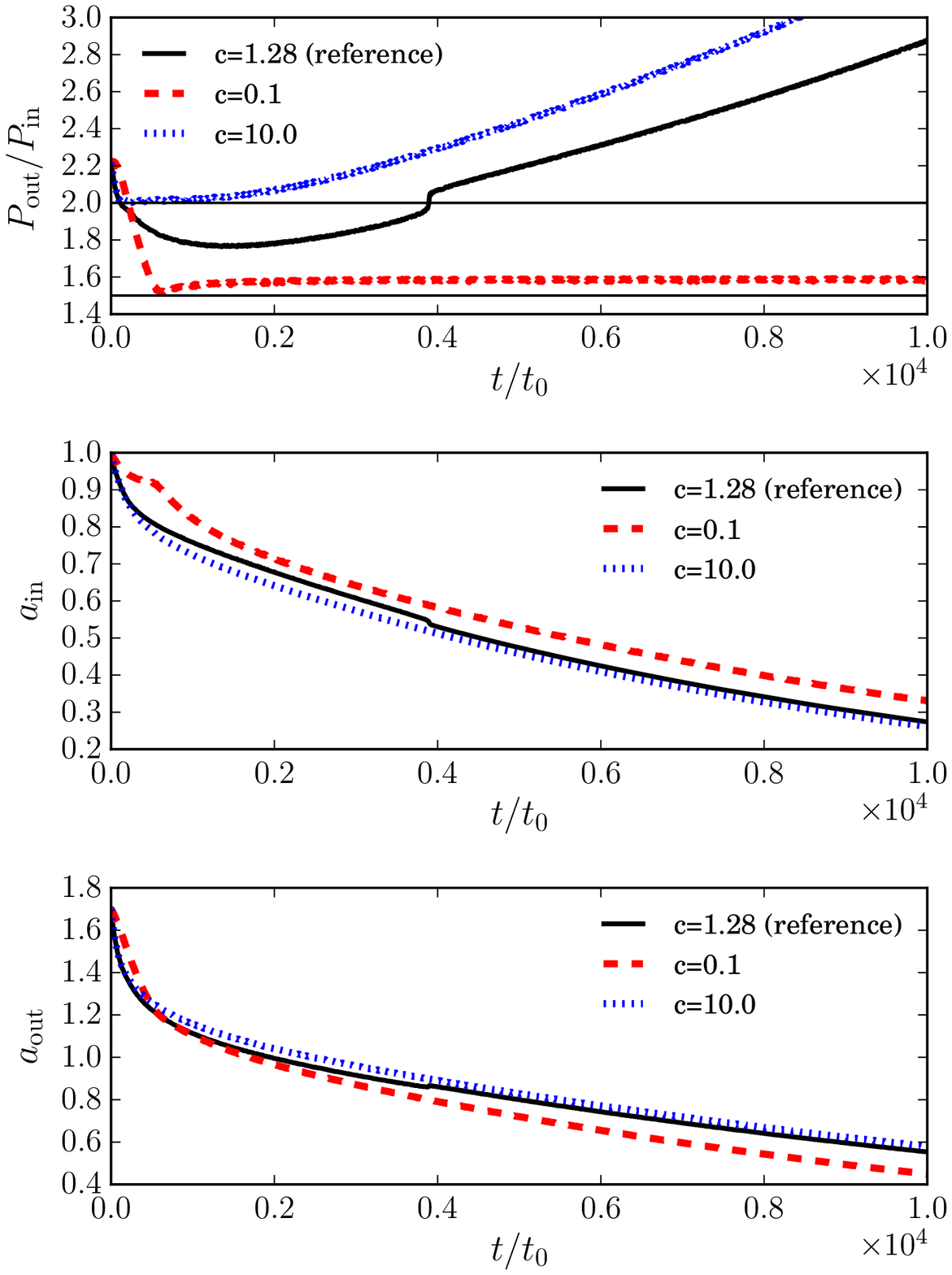}}
		\caption{
		The time variations of the period ratio (upper panel), the semi-major-axis of the inner planet (middle panel) and the 
		semi-major axis of the outer 
		planet (lower panel), in the case of $\mpin/\mstar=3\times 10^{-4}$, $\mpout/\mstar=5\times 10^{-4}$, $h/R=0.05$, and $\alpha=10^{-3}$.
		The solid, dashed and dotted lines denote the cases of $c=1.28$ (reference), $c=0.1$, and $c=10.0$, respectively.
		\label{fig:porbits_cvar}
		}
	\end{center}
\end{figure}
We have carried out the three-body simulations with three different values of $c$, that is, $c=1.28$ (the reference case), $c=10.0$ 
(the case of a long damping timescale), and $c=0.1$ (the case of a short damping timescale).
Note that in Section~\ref{sec:3body_sims}, we have adopted $c=1.28$.
In Figure~\ref{fig:porbits_cvar}, we show the time variations of the period ratio and the semi-major axes of the inner and outer planets 
when $\mpin/\mstar=3\times 10^{-4}$, $\mpout/\mstar=5\times 10^{-4}$, $h/R=0.05$, and $\alpha=10^{-3}$.
In the case of $c=1.28$, the period ratio decreases with time until $t\simeq 2000\ t_0$, and after that, it starts to increase with time and continues like this until the end of the calculations.
When the value of $c$ is much larger than the reference value (i.e., $c=10.0$), the time variation of the period ratio is similar to that seen already in the reference case, though the value of the period ratio at the turnover is a bit different.
The time variations of the semi-major axes of the inner and outer planets are also similar to those in the reference case.
On the other hand, if the value of $c$ is much smaller than the reference value (i.e., $c=0.1$), the planet pair is captured into the 3:2 mean-motion resonance.
Hence, the outcome here is different from the cases with $c=1.28$ and $0.1$.
In the case with $c=0.1$, the planetary migration is affected by the strong damping of the eccentricity, and therefore, the time variations of the semi-major axes of the inner and outer planets are different from those obtained for the smaller values of $c$.
According to the previous studies \citep{Goldreich_Sari2003,Duffell_Chiang2015}, when the planet forms a deep gap, the disk--planet interactions work 
to excite the eccentricity of the gap-opening planet, rather than damping the eccentricity.
This means that at least for massive planets which are able to open such a deep gap in the disk, a long damping timescale of the eccentricity (and the large value of $c$) may be appropriate.

\begin{figure}
	\begin{center}
		\resizebox{0.48\textwidth}{!}{\includegraphics{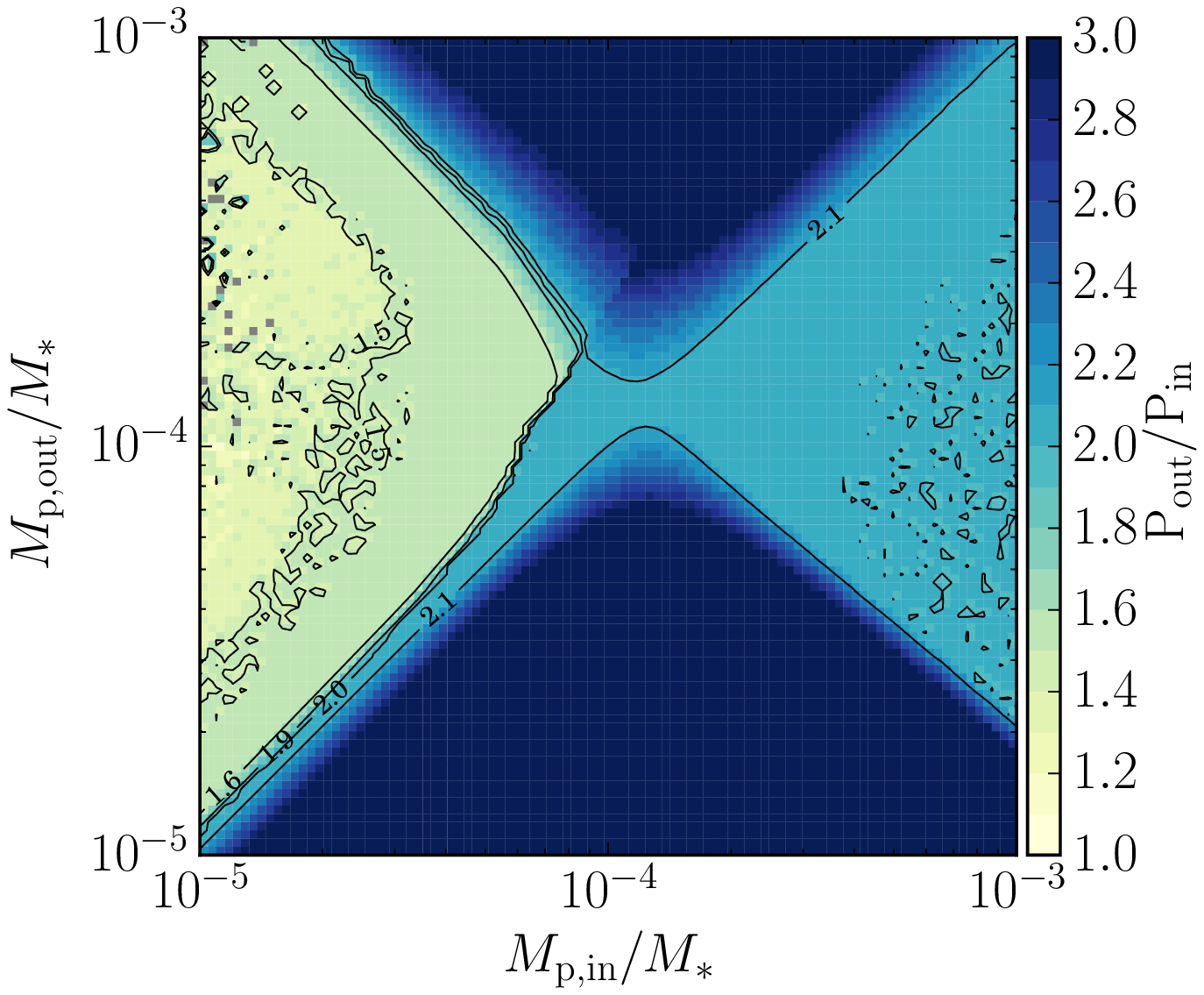}}
		\resizebox{0.48\textwidth}{!}{\includegraphics{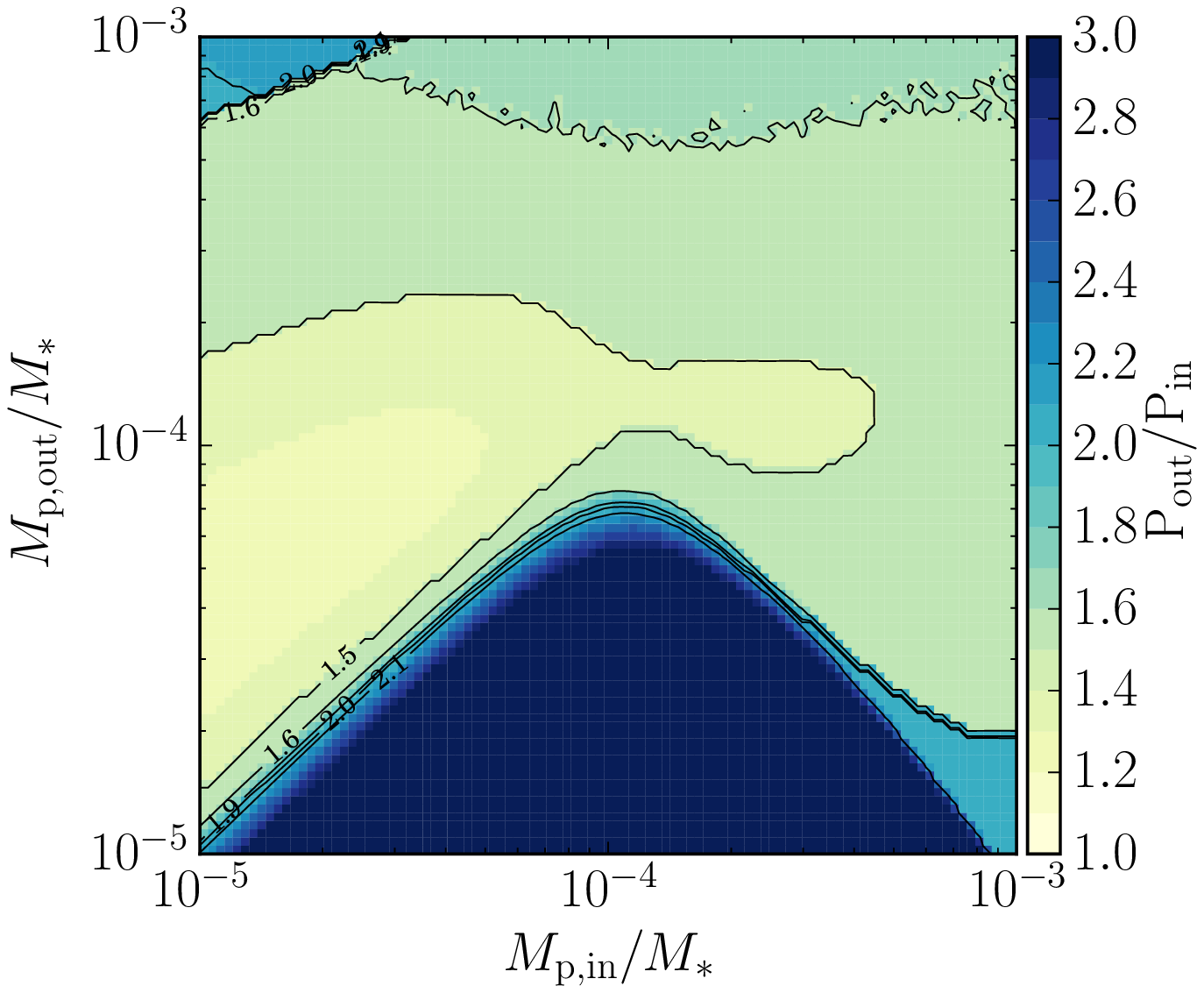}}
		\caption{
		The same as in Figure~\ref{fig:period_ratio_h0.05_a1e-3}, but $c=10.0$ in the left panel and $c=0.1$ in the right panel.
		\label{fig:period_ratio_h0.05_a1e-3_cvar}
		}
	\end{center}
\end{figure}
In Figure~\ref{fig:period_ratio_h0.05_a1e-3_cvar}, we show the period ratios of the planet pairs at the end of the three-body simulations adopting $c=10.0$ and $0.1$.
The other parameters (i.e., masses of the planets and values of $h/R$ and $\alpha$) are the same as in the calculations illustrated in Figure~\ref{fig:period_ratio_h0.05_a1e-3}.
In the case of $c=10.0$, the distribution of the period ratio is quite similar to that shown in Figure~\ref{fig:period_ratio_h0.05_a1e-3} 
(in the case of $c=1.28$), though in some cases, the lighter inner planet is strongly scattered by the heavier outer planet.
On the other hand, if $c=0.1$, the planet pairs are more likely captured into the 3:2 mean-motion resonance when the outer planet is larger 
than $\mpout/\mstar \simeq 1\times 10^{-4}$, as compared with the case of Figure~\ref{fig:period_ratio_h0.05_a1e-3}.
Hence, our results on the period ratios are not significantly affected by the damping timescale of the eccentricity, unless the damping timescale is very short.

\section{Critical timescale for 2:1 mean-motion resonance} 
\label{sec:tcrit}
Here we discuss the critical timescale for 2:1 mean-motion resonance capture.
For simplicity, we fix the mass of the outer planet as $\mpout/\mstar=3.2\times 10^{-4}$.
For the mass of the inner planet, we adopt $\mpin/\mstar=2 \times 10^{-5}$ and thus $\mpout/\mpin > 1$ in this case.
As a comparison, we also carried out the simulation with $\mpin/\mstar=8\times 10^{-4}$.
In the comparison case, $\mpout/\mpin<1$ as in \cite{Ogihara_Kobayashi2013} and the migration velocity of the inner planet given by Equation~(\ref{eq:tmig_timevar}) is similar to that in the case of $\mpin/\mstar=2\times 10^{-5}$ ($\tmig \simeq 1.3\times 10^{5} \omegakp^{-1}$ when $\sigmazero=3\times 10^{-4}$).
The disk parameters are the same as those in the case shown in Figure~\ref{fig:period_ratio_h0.05_a1e-3}, except $\sigmazero$.
By changing $\sigmazero$, we can look for the critical timescale for the 2:1 mean-motion resonance (the migration is faster as $\sigmazero$ increases as can be seen in Equation~\ref{eq:tmig_timevar}).
\begin{figure*}
	\begin{center}
		\resizebox{0.98\textwidth}{!}{\includegraphics{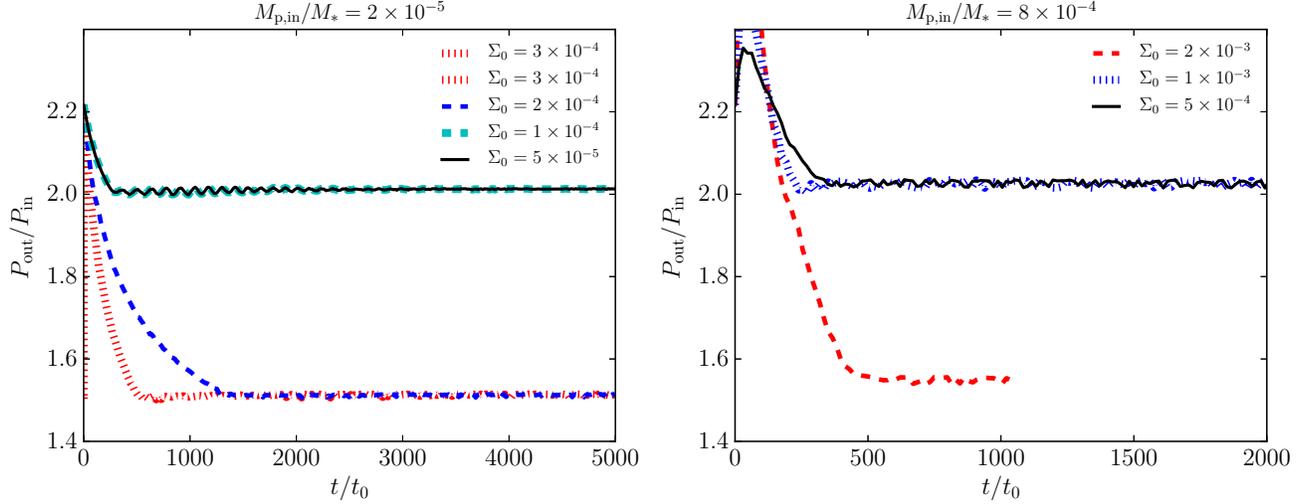}}
		\caption{
		Time variations of a period ratio for various $\sigmazero$.
		In the left panel, $\mpin/\mstar=2\times 10^{-5}$ and in the right panel, $\mpin/\mstar=8\times 10^{-4}$.
		The mass of the outer planet is $\mpout/\mstar=3.2\times 10^{-4}$ in both panels.
		Other parameters are the same as those in the case shown in Figure~\ref{fig:period_ratio_h0.05_a1e-3}.
		We terminated the simulations when the inner planet reaches $0.3R_0$.
		\label{fig:critical_timescale}
		}
	\end{center}
\end{figure*}
Figure~\ref{fig:critical_timescale} shows the time variations of the period ratio for various $\sigmazero$.
In the left panel of the figure ($\mpin/\mstar=2\times 10^{-5}$), the planet pair is captured in the 2:1 mean-motion resonance when $\sigmazero<10^{-4}$, and hence the critical timescale is estimated as $\tmigin-\tmigout \simeq 2.3\times 10^{5}\omegakp^{-1}$, which is consistent with that given by Equation~(\ref{eq:tcrit}) (it is $4\times 10^{5}\omegakp^{-1}$).
In the right panel of Figure~\ref{fig:critical_timescale}, the threshold of the surface density for 2:1 mean-motion resonance capture is about $\sigmazero=10^{-3}$.
In this case, the critical timescale is estimated by $\tmigin-\tmigout \simeq 10^{4}\omegakp^{-1}$.
Equation~(\ref{eq:tcrit}) gives $4\times 10^{3}$, which is consistent with our value within the factor of $2$ -- $3$.

As shown above, the critical timescale is shorter as $\mpin$ increases and Equation~(\ref{eq:tcrit}) can be applicable even when $\mpout/\mpin>1$.
As the mass of the inner planet decreases, the critical timescale becomes longer.
That is, the planet pair easily passes the 2:1 mean-motion resonance when the inner planet is small.
As discussed in Section~\ref{subsec:resonant_capture}, this tendency can explain the result shown in Figure~\ref{fig:period_ratio_h0.05_a1e-3}, and the planet pair evolving convergently is likely to be captured into 3:2 (or those with larger integers, as 4:3 for example) mean-motion resonance when $\mpin < \mptrans$.
On the other hand, the most of the pairs are captured into the 2:1 mean-motion resonance when $\mpin>\mptrans$ because the critical timescale is very short due to a massive inner planet.

\section{List of two-planet systems}
 \label{sec:list_twoplanet_sys}
The planet pairs shown in Figure~\ref{fig:massratio_vs_pratio_comp_with_obs} are listed in Table~\ref{tab:2planet_sys}.
\begin{deluxetable}{ccccccccc}
\tablecaption{List of two-planet systems in Figure~\ref{fig:massratio_vs_pratio_comp_with_obs} \label{tab:2planet_sys}}
\tablewidth{0pt}
\tablehead{
\colhead{Name} & \colhead{$\pin$ (day)} & \colhead{$\pout$ (day)} & \colhead{$\mpin$ ($M_{\oplus}$)} & \colhead{$\mpout$ ($M_{\oplus}$)} & \colhead{$M_{\ast} (M_{\odot})$} & \colhead{$\pout/\pin$} & \colhead{$\mpout/\mpin$} & Category\tablenotemark{a}
} 
\startdata
HD 1461 &5.77   &13.51  &6.75   &5.86   &1.02   &2.34   &0.87   &1\\
KOI-1599 (Kepler-1659)        &13.61  &20.44  &9.44   &4.82   &1.02   &1.50   &0.51   &2\\
Kepler-87       &114.74 &191.23 &340.00 &6.67   &1.10   &1.67   &0.02   &2\\
Kepler-36       &13.84  &16.24  &4.67   &8.33   &1.07   &1.17   &1.79   &3\\
24 Sex  &452.80 &883.00 &663.33 &286.67 &1.54   &1.95   &0.43   &4\\
7 CMa   &735.10 &996.00 &616.67 &290.00 &1.34   &1.35   &0.47   &4\\
HD 106315       &9.55   &21.06  &13.21  &15.94  &1.09   &2.20   &1.21   &4\\
HD 113538       &663.20 &1818.00        &120.00 &310.00 &0.58   &2.74   &2.58   &4\\
HD 128311       &453.02 &921.54 &589.67 &1263.00        &0.83   &2.03   &2.14   &4\\
HD 155358       &194.30 &391.90 &330.00 &273.33 &0.92   &2.02   &0.83   &4\\
HD 176986       &6.49   &16.82  &6.02   &9.63   &0.79   &2.59   &1.60   &4\\
HD 20003        &11.85  &33.92  &12.23  &15.13  &0.88   &2.86   &1.24   &4\\
HD 200964       &606.30 &852.50 &533.00 &404.67 &1.39   &1.41   &0.76   &4\\
HD 202696       &517.80 &946.60 &665.33 &621.33 &1.91   &1.83   &0.93   &4\\
HD 21693        &22.68  &53.74  &8.63   &18.23  &0.80   &2.37   &2.11   &4\\
HD 23472        &17.67  &29.62  &18.79  &18.02  &0.75   &1.68   &0.96   &4\\
HD 33844        &551.40 &916.00 &670.00 &583.33 &1.78   &1.66   &0.87   &4\\
HD 45184        &5.89   &13.14  &12.80  &9.23   &1.03   &2.23   &0.72   &4\\
HD 47366        &359.15 &682.85 &766.67 &626.67 &2.19   &1.90   &0.82   &4\\
HD 5319 &637.10 &872.20 &518.67 &351.00 &1.27   &1.37   &0.68   &4\\
HD 60532        &201.90 &600.10 &353.33 &836.67 &1.50   &2.97   &2.37   &4\\
HD 73526        &188.30 &379.10 &1026.67        &750.00 &1.01   &2.01   &0.73   &4\\
HIP 54373       &7.76   &15.14  &9.04   &13.05  &0.57   &1.95   &1.44   &4\\
HIP 65407       &28.12  &67.30  &142.67 &261.33 &0.93   &2.39   &1.83   &4\\
Kepler-117      &18.80  &50.79  &31.33  &613.33 &1.13   &2.70   &19.57  &4\\
OGLE-2006-BLG-109L      &1788.50        &4927.50        &242.33 &90.00  &0.51   &2.76   &0.37   &4\\
TOI-216 &17.09  &34.56  &31.46  &209.76 &0.87   &2.02   &6.67   &4\\
TYC 1422-614-1  &198.40 &559.30 &833.33 &3333.33        &1.15   &2.82   &4.00   &4\\
Teegarden's Star        &4.91   &11.41  &1.10   &1.16   &0.09   &2.32   &1.06   &4\\
gam Lib &415.20 &964.60 &340.00 &1526.67        &1.47   &2.32   &4.49   &4\\
rho CrB &39.85  &102.54 &348.30 &26.22  &0.89   &2.57   &0.08   &4\\
\enddata
\tablenotetext{a}{1:the case of $\mpin,\mpout < \mptrans$, 2: the case of $\mpin > \mptrans, \mpout < \mptrans$, 3: the case of $\mpin<\mptrans, \mpout > \mptrans$, 4: the case of $\mpin,\mpout > \mptrans$}
\end{deluxetable}



\end{document}